\documentclass[
    aps,
    prx,
    reprint,
    amsmath,
    amssymb,
    superscriptaddress,
    longbibliography
]{revtex4-2}
\usepackage[T1]{fontenc}
\usepackage{graphicx}
\graphicspath{{figs/}}
\usepackage{xcolor}
\usepackage{booktabs}
\usepackage[french,english]{babel}
\usepackage{hyperref}
\hypersetup{
    colorlinks=true,
    linkcolor=blue,
    urlcolor=brown,
    citecolor=purple,
    pdftitle={Designing turbulence through hierarchies}
}

\begin{document}

\title{Designing turbulence through hierarchies}

\author{Flavio Tuteri}
\thanks{These authors contributed equally}
\affiliation{Laboratoire de Physique de l'\'Ecole normale sup\'erieure, ENS, Universit\'e PSL, CNRS, Sorbonne Universit\'e, Universit\'e Paris Cit\'e, Paris, F-75005, France}
\email{flavio.tuteri@phys.ens.psl.eu}
\author{André Freitas}
\thanks{These authors contributed equally}
\affiliation{Department of Physics and INFN, University of Rome “Tor Vergata”, Rome, Italy}
\email{andre.freitas@roma2.infn.it}
\affiliation{LTCI, Télécom Paris, IP Paris, Palaiseau, France}

\date{\today}

\begin{abstract}
Turbulence is defined by a set of emergent and singular phenomena: a
constant-flux energy cascade, intermittency, a dissipation rate that
survives the vanishing of viscosity, and the spontaneous loss of
determinism of the inviscid limit. Which physical ingredients produce them
remains unclear, because in fluids they always come together with
incompressibility, spatial structure and vortex dynamics. Here we show that
a hierarchy of inertia, coupled by the purely geometric
nonlinearity of rigid constraints, is enough. A planar chain of rigid
pendula whose lengths and masses decrease geometrically, coupled only
through nearest-neighbor constraints, develops a far-from-equilibrium
cascade that replaces thermal equipartition, with statistically local and
time-irreversible energy transfer despite intrinsically nonlocal coupling.
Its fluctuations are intermittent, yet self-similarity is restored by a
hidden scale symmetry. As the viscosity vanishes, the dissipation rate
saturates at a finite value, and thermal noise consistent with
fluctuation--dissipation balance is amplified from the smallest to the
largest scales within a fraction of a large-scale period. Because the
hierarchy can be tuned and the same dynamics can be realized mechanically
or electrically, turbulence becomes a property that can be designed and
measured in the laboratory.
\end{abstract}

\maketitle

\section{Introduction}
\label{sec:intro}

Turbulence is the paradigm of emergent behavior far from equilibrium.
Energy injected into a flow at large scales is handed down by the
nonlinearity to progressively smaller scales, until viscosity turns it into
heat. Across the intermediate inertial range the mean energy flux does not
depend on scale, the cornerstone of Kolmogorov's phenomenology
\cite{kolmogorov1941,obukhov1941,frisch1995turbulence,alexakis2018cascades}.
This mean cascade is only the beginning. Turbulence is defined by a set of
emergent properties that have no counterpart in laminar flows or in
systems at thermal equilibrium, and several of them are \emph{singular}:
they describe what happens as the viscosity, the only regularization of the
dynamics, is sent to zero.

The first is intermittency. High-order statistics depart systematically
from the dimensional self-similarity predicted by a constant flux
\cite{anselmet1984,meneveau1991}, with anomalous exponents that motivated
multifractal and hierarchical descriptions of the cascade
\cite{mandelbrot1974,meneveau1987,sheleveque1994,benzi1993ess}. The
second is the dissipative anomaly, or zeroth law of turbulence: the
dissipation rate remains finite as the viscosity vanishes
\cite{onsager1949,zeroth_law}. Whether this anomaly is strong, with a
strictly positive limit, or weak, with a dissipation that vanishes slowly,
is still debated in three-dimensional turbulence \cite{zeroth_law}. The
third is spontaneous stochasticity
\cite{falkovich2001,mailybaev2016,eyink2025chaosfluctuationsanomaliesspontaneous}.
In the joint limit of vanishing viscosity and vanishing microscopic noise,
solutions starting from the same initial condition need not converge to a
unique trajectory: the limiting distribution remains broad at any positive
time, even though it is a delta function initially. This is fundamentally
different from deterministic chaos, in which the uncertainty at a fixed
time vanishes together with the initial perturbation. The idea goes back
to Lorenz's observation that, in a system with many scales, errors climb
the hierarchy in a finite time \cite{lorenz1969,leithkraichnan1972}.
Fluctuating-hydrodynamics
studies showed that even thermal fluctuations originating in the dissipation
range can reach the energy-containing scales within a few eddy-turnover
times \cite{ruelle1979,bandak2022,bandak2024,freitas2026stochasticity,ortizbernal2026}.

In fluids, these phenomena always come together with incompressibility,
nonlocal pressure, vortex stretching and spatial structure. It is then
hard to tell which ingredients are responsible for which property, and
which are merely incidental. Reduced models have been instrumental in
separating them. Shell models replace the
velocity field by a single hierarchy of scale variables with couplings
local in scale, and reproduce the cascade, intermittency and singular
dynamics \cite{gledzer1973,goy,sabra,shellreview,jensen1991,kadanoff1995,mailybaev2013blowup}.
Their anomalous statistics have recently been traced to hidden scale
symmetries, which restore self-similarity in dynamically rescaled variables
\cite{mailybaev2021hidden,mailybaev2022hidden,mailybaev2022spatiotemporal,mailybaev2023hidden},
and which have since been identified in Navier--Stokes turbulence, in
passive scalars and in large-eddy simulations
\cite{mailybaevthalabard2022,calascibetta2025hidden,magacho2025hidden}.
Multi-branch shell models add a growing number of degrees of freedom per
scale and allow spatial organization and branching to be studied in the
same framework
\cite{tuteri2026fixedpoints,tuteri2026multibranch2d,tuteri2026intermittency}.

Shell models, however, are designed rather than derived. Their couplings
are chosen to be local in scale and to conserve the relevant invariants,
and they do not describe a physical system that could be built and
measured. This leaves two basic questions open. First, can the full
turbulent phenomenology emerge in a genuine physical system whose only
multiscale ingredient is a hierarchy, and whose couplings are dictated by
Newton's laws rather than chosen by the modeler? Second, if it can, can
the resulting turbulence be engineered by choosing the hierarchy?

\begin{figure*}
    \centering
    \includegraphics[width=\textwidth]{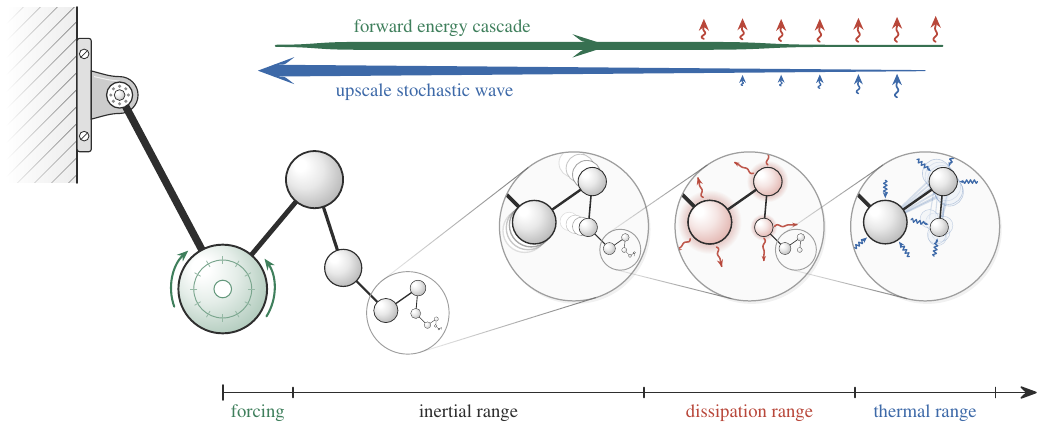}
    \caption{\textbf{The hierarchical pendulum chain.}
    A planar chain of point masses connected by massless rigid rods is suspended from a fixed pivot.
    Rod lengths and masses decrease geometrically along the chain, e.g., $\ell_{k+1}=\ell_k/2$ and $m_{k+1}=m_k/2$ (the aspect ratio is adjusted for visualization).
    The bubbles show successive magnifications of the tail, each reproducing the entire chain and illustrating the discrete scale symmetry imposed by the architecture.
    The dominant physical mechanisms vary along the hierarchy, as summarized by the link-index axis $k$ at the bottom.
    Energy is injected at the largest scales by antisymmetric torques $\tau_2=-\tau_1$ acting on the first two links, equivalent to a motor at their joint driving them in opposite directions (green).
    The turbulent state is characterized by a constant mean energy flux toward smaller scales across the inertial range, where the links evolve under conservative rod tensions (first bubble).
    In the dissipative range, energy is removed by the bending viscosity $\eta$, which damps relative angular motion between neighboring links (red, second bubble).
    Fluctuation–dissipation-consistent thermal torques can be introduced at the same small scales (blue): weak where viscous effects dominate, their influence becomes strongest at the end of the chain, in the thermal range (third bubble).
    Because the dynamics is strongly nonlinear over many scales, with inertia dominating dissipation across the inertial range and turnover times that shorten geometrically along the hierarchy, microscopic uncertainty is amplified and propagates upscale through the hierarchy \cite{lorenz1969}, forming a stochastic wave, a hallmark of spontaneous stochasticity \cite{mailybaev2016,bandak2024}.
    The two arrows at the top schematically illustrate the downscale energy flux and the upscale propagation of the uncertainty saturation front, which reaches each scale on a timescale comparable to its characteristic local turnover time.
    }
    \label{fig:setup}
\end{figure*}

In this paper we answer both questions in the affirmative.
We study the cascading pendulum, a planar chain of rigid links whose
lengths and masses decrease geometrically (Fig.~\ref{fig:setup}),
introduced by Bandak in his doctoral thesis~\cite{bandakthesis} as an
experimentally realizable model of spontaneous stochasticity. The thesis
identified the geodesic structure and scale symmetries of the chain and,
in simulations with dissipation acting only at the last joint, an extreme
form of hyperviscous
regularization~\cite{frisch2008hyperviscosity,agrawal2020hyperviscous},
gave first indications of a constant flux, spontaneous stochasticity and
scale-invariant multipliers. Whether the chain sustains a turbulent
stationary state, displays anomalous scaling and exhibits a dissipative
anomaly remained open. Here we address these questions under controlled
conditions. We let the dissipative scale emerge from a bending viscosity
acting at every joint, so that the inertial range is separated from the
end of the chain, and vary the viscosity over ten decades at a fixed chain
length that well resolves the dissipative range at all viscosities; we add
the thermal noise prescribed by fluctuation--dissipation balance, derive
the exact energy flux through each link, and quantify the uncertainties of
all statistics. This allows us to establish the turbulent phenomenology of
the chain, to address properties not considered before, and to show that
the hierarchy itself is a control parameter.

Each link is constrained only by its
neighbors, yet inertia couples every link to the whole chain: the motion
of each mass contributes to the inertia carried by all larger links. The
interaction is thus nonlocal in scale, unlike in shell models, and the
constraint tension, which enforces inextensibility, plays a role analogous
to pressure in incompressible flows. The chain has no potential energy:
its nonlinearity is purely geometric, arising from the
configuration-dependent inertia imposed by the rigid links. 
Its Lagrangian depends only on
an inertia matrix, and can equally be realized by an electromechanical
circuit (Appendix~\ref{app:experimental}).
Mechanically,
it sits between the free rotation of a rigid body and the dynamics of an
ideal fluid, all three being geodesic motions in Arnold's geometric
picture \cite{arnold1966}, with an explicit hierarchy of scales added.
We drive the chain at its largest scales. The chain then reproduces the
full phenomenology of turbulence, in three stages that structure this paper.
(i)~\emph{The cascade.} In place
of thermal equipartition, the chain sustains a constant-flux cascade far
from equilibrium. The transfer is statistically local, because
instantaneous exchanges with remote scales cancel in the mean, and time
irreversible, even when the only irreversible term in the equations
vanishes. (ii)~\emph{Its fluctuations.} The cascade is intermittent, with
anomalous exponents carried by the fluctuations of the exact mechanical
flux, and its self-similarity is restored by a hidden scale symmetry.
Thermal noise thermalizes only the smallest scales and leaves the cascade
intact. (iii)~\emph{Its singular limit.} As the viscosity is reduced over
ten decades, the dissipation rate saturates at a finite value, and
microscopic noise is amplified and climbs from the smallest to the largest
scales, against the energy cascade, within a fraction of a large-scale
period. Finally, because the mass and length ratios set the dimensional
scaling of the cascade, the hierarchy itself becomes a design parameter
for turbulence.

Section~\ref{sec:model} introduces the chain, its exact energy flux, the
forcing, dissipation and noise, and the dimensional predictions that
distinguish a cascade from equipartition and make the hierarchy a control
parameter. Section~\ref{sec:results} presents the three stages above.
Section~\ref{sec:discussion} discusses what the results imply for the
ingredients of turbulence, for the design of turbulent systems, and for
experiments. Derivations and numerical methods are collected in the
Appendices.

\section{Hierarchical architectures}
\label{sec:model}

Nonlinear interactions in complex systems can give rise to \emph{emergent} organization.
For example, in three-dimensional turbulence, this self-organization
manifests itself as a hierarchy of interacting scales, with smaller
structures evolving at progressively higher frequencies
\cite{kolmogorov1941,frisch1995turbulence}.
When the dynamics is described in terms of the appropriate intrinsic clock of each scale, a statistical scale invariance emerges in the turbulent state \cite{mailybaev2025multitime}.
Here, we reverse this perspective: rather than allowing a hierarchy
to emerge from the dynamics, we encode it directly into the architecture
of a nonlinear Lagrangian system. We then investigate how turbulent
behavior arises from the interplay between deterministic chaos and
\emph{imposed} scale symmetry.
A natural starting point is a chain of coupled pendula.
Already at two degrees of freedom, the double pendulum exhibits
deterministic chaos \cite{shinbrot1992}. Extending this elementary system across
geometrically separated scales provides a minimal setting in which
to explore the collective dynamics of a nonlinear hierarchy.

\subsection{Pendulum chain}
\label{sec:chain}

Our starting point is a planar chain of $N$ point masses $m_k$, connected by massless rigid rods of lengths $\ell_k$, as illustrated in Fig.~\ref{fig:setup}.
In Cartesian coordinates, the state of the chain is specified by the positions $\mathbf r_k$ of the masses and their velocities.
The chain is suspended from a fixed anchor at $\mathbf r_0$.
A hierarchy of timescales is established by geometrically scaling the lengths and masses:
\begin{equation}
\ell_{k+1}=\ell_k/\gamma,\qquad m_{k+1}=m_k/\gamma,
\qquad \gamma>1,
\label{eq:hierarchy}
\end{equation}
with prescribed $\ell_1$ and $m_1$ (numerical simulations are performed for $\gamma=2$).
Gravity is neglected in the model, as its influence is expected to be confined to the largest scales of the chain.
The Lagrangian is therefore purely kinetic, subject to the rigidity constraints of the links:
\begin{equation}
    \label{eq:lagr1}
    \mathcal{L}=\frac{1}{2}\sum_{k=1}^N
    \Big\{m_k
    \lvert\dot{\mathbf r}_k\rvert^2
    +
    \lambda_k
    \big(
    \lvert\mathbf r_{k}-\mathbf r_{k-1}\rvert^2
    -\ell_k^2
    \big)
    \Big\},
\end{equation}
where $\lambda_k$ are Lagrange multipliers enforcing the inextensibility of the rods, physically realized through tension forces.
There are no field-mediated interactions: in Cartesian coordinates, the system has a diagonal kinetic energy, and all interactions arise from the geometric constraints.
The simplicity of the diagonal kinetic energy comes at the cost of evolving the dynamics on a constrained manifold.
To avoid explicitly projecting the evolution onto the constraint manifold, the trick is to use a parametrization that automatically satisfies the rigidity constraints.
Each point mass can only rotate around the preceding one in the hierarchy, so the configuration can be fully described by angles $\theta_k$ measured from a fixed direction.
The price of eliminating the explicit constraints is nonlocality:
\begin{equation}
    \mathbf r_k=\mathbf r_0+\sum_{j=1}^{k}\ell_j(\sin\theta_j,\,-\cos\theta_j).
    \label{eq:coordinates}
\end{equation}
The position of a point mass along the hierarchy can thus be reconstructed only by descending the chain from the anchor to that point.
Behind this apparent increase in complexity lies the unavoidable contribution of the tension forces, which couple the motion along the entire chain to enforce rigidity at every instant.
The phase space is then described by $(\boldsymbol{\theta},\boldsymbol{\omega})$, collecting the $N$ angular positions and their corresponding angular velocities.
Substituting the parametrization \eqref{eq:coordinates} into the Lagrangian \eqref{eq:lagr1}, we obtain the purely kinetic quadratic form
\begin{equation}
\label{eq:lagr2}
\mathcal{L}=\frac{1}{2}\boldsymbol{\omega}\cdot\mathcal{A}(\boldsymbol{\theta})\boldsymbol{\omega},
\end{equation}
with the configuration-dependent inertia matrix
\begin{equation}
\mathcal A_{kj}(\boldsymbol\theta)=M_{kj}\ell_k\ell_j\cos(\theta_j-\theta_k),
\label{eq:metric}
\end{equation}
where $M_{kj}=\sum_{i\ge\max(k,j)} m_i$ is the total mass carried by the chain beyond the outermost of links $k$ and $j$.
This is the archetypal structure studied in this work: a purely kinetic Lagrangian with a configuration-dependent energy metric possessing a scaling property, inherited in this case from the geometric scaling of masses and lengths \eqref{eq:hierarchy}.
Lagrangians with the same structure reproduce the dynamics numerically investigated here, enabling the design of physical systems that exhibit the same phenomenology but rely on interactions that are easier to control and implement in the laboratory.
In Appendix~\ref{app:experimental}, we present an electromechanical realization, showing that gravity can be eliminated altogether in a physical implementation.
Except for a set of configurations of measure zero, the coupling matrix \eqref{eq:metric} is fully dense, meaning that every pair of angular degrees of freedom is coupled, with interaction strengths scaling exponentially along the hierarchy of inertia.
Each link carries the inertia of all the masses it supports: the interaction is therefore fully nonlocal in scale, even though it arises from nearest-neighbor constraints.
In this respect, the chain resembles incompressible flows, where pressure acts as a Lagrange multiplier enforcing incompressibility and couples distant scales instantaneously, rather than shell models of turbulence, where interactions are restricted by construction to a few neighboring shells (quasi-locality) \cite{goy,sabra,shellreview}.
Likewise, the tension forces in the chain enforce rigidity and mediate nonlocal coupling involving $O(N^2)$ terms.
Nevertheless, this nonlocal system remains computationally accessible: the coupling matrix \eqref{eq:metric} has a rank-two semiseparable structure that enables exact $O(N)$ linear algebra (Appendix~\ref{app:numerics}), so that simulations scale like those of quasi-local models.
The Euler–Lagrange equations governing the conservative dynamics read
\begin{equation}
\mathcal A(\boldsymbol\theta)\dot{\boldsymbol\omega}=\boldsymbol b(\boldsymbol\theta,\boldsymbol\omega),
\label{eq:conservative-compact}
\end{equation}
where the right-hand side consists of the centrifugal nonlinear terms
\begin{equation}
b_k=\sum_{j=1}^N M_{kj}\ell_k\ell_j\sin(\theta_j-\theta_k)\omega_j^2.
\label{eq:centrifugal}
\end{equation}
These terms reveal a distinctive feature of the dynamics: they are quadratic in the angular velocities, with no contribution linear in the angular displacements.
Consequently, reducing the kinetic energy does not lead to a regime of small oscillations governed by fixed natural frequencies.
Instead, the characteristic frequencies are determined by the motion itself: rescaling all angular velocities simply rescales the rate at which the same trajectory is traversed.
As for the Euler equations of an ideal fluid, this absence of an intrinsic time scale reflects the geodesic character of the dynamics noted in the Introduction. The continuum limit shares it: a chain of point masses on rigid rods becomes an inextensible string, or whip, whose motion is again a geodesic flow and whose tension, fixed instantaneously by the motion, plays the role of pressure \cite{preston2011geometry,preston2011whips}. Since $m_k/\ell_k$ is constant, the chain studied here is a lumped discretization of a uniform whip, refined geometrically toward its free end.

\subsection{Symmetries and conservation laws}
\label{sec:energy-budget}

The conservative dynamics exhibits symmetries of the equations of motion, some of which also leave the action functional invariant and yield conserved quantities through Noether’s theorem.
The time-rescaling symmetry
\begin{equation}
\theta_k(t)\mapsto\theta_k(t/a),
\label{eq:time_scaling}
\end{equation}
is a symmetry of the equations of motion \eqref{eq:conservative-compact} but not of the action.
This symmetry provides the basis for dimensional arguments leading to Kolmogorov-like turbulent scaling predictions \cite{kolmogorov1941}.

The geometric hierarchy \eqref{eq:hierarchy} introduces a discrete scale symmetry. For a chain extending infinitely in both directions, the equations of motion are invariant under a shift of the link index,
\begin{equation}
\theta_k(t)\mapsto\theta_{k+1}(t).
\label{eq:scale_shift}
\end{equation}
Unlike in the Euler equations, where spatial and temporal rescalings are linked by the scaling of the velocity field, here time rescaling and scale shifts act independently, because spatial scales are encoded directly in the degrees of freedom.

For a finite chain, the conservative dynamics possesses two conserved quantities: the total energy $E$ and the total angular momentum $J$ about the fixed pivot. They follow, respectively, from time-translation invariance and global rotational invariance of the action. The latter requires the absence of gravity, so that no direction in the plane is preferred.
Both quantities are sums of contributions from individual masses,
\begin{equation}
\label{eq:local_observables}
    e_k=\frac{m_k}{2}\lvert\dot{\mathbf r}_k\rvert^2,
    \qquad
    j_k=m_k(\mathbf r_k-\mathbf r_0)\wedge\dot{\mathbf r}_k,
\end{equation}
such that $E=\sum_k e_k$ and $J=\sum_k j_k$.
Although the total energy is conserved, the individual contributions $e_k$ change as energy is transferred along the chain. Remarkably, their evolution takes the form of a local energy budget,
\begin{equation}
    \frac{de_k}{dt}=\Pi_k-\Pi_{k+1},
    \quad
    \Pi_k=\lambda_k\dot{\mathbf r}_k\cdot(\mathbf r_k-\mathbf r_{k-1}),
    \label{eq:local-budget}
\end{equation}
where the two terms differ only by a shift along the hierarchy, naturally identifying $\Pi_k$ and $\Pi_{k+1}$ as the incoming and outgoing energy fluxes at scale $k$.
At the boundaries, $\Pi_1=0$ because the pivot is fixed, while $\Pi_{N+1}=0$ because the chain terminates at mass $N$ (Appendix~\ref{app:flux}).
The telescoping structure makes the interpretation of $\Pi_k$ explicit:
$d_t\sum_{i<k}e_i=-\Pi_k$ is the rate of energy transfer through link $k$, with $\Pi_k>0$ corresponding to a forward transfer from larger to smaller scales.
The energy flux is a physical observable: it is the rate of work transmitted by a constraint force and can be expressed in terms of the angular coordinates and velocities. It is therefore directly accessible in a laboratory realization.
Equation~\eqref{eq:local-budget} coincides on average with the flux defined in
Ref.~\cite{bandakthesis} as the work done on the subchain in the frame of its
suspension point, but the two differ instantaneously by a total time
derivative, and only Eq.~\eqref{eq:local-budget} provides an exact local budget.

\subsection{Forcing, dissipation and noise}
\label{sec:forcing-dissipation}

To sustain a nonequilibrium statistically stationary state, the conservative dynamics must be supplemented by a source and a sink of energy.
In a mechanical realization, energy is supplied by a motor and removed by friction.

Energy is injected through an antisymmetric torque acting on the first two links,
\begin{equation}
\label{eq:forcing}
\tau_k=\tau(t,\boldsymbol{\omega})(\delta_{k1}-\delta_{k2}),
\end{equation}
which drives their relative angular motion without exerting a net torque on the chain, thereby preserving the total angular momentum.
The forcing amplitude may depend on the instantaneous state of the system, as is common in numerical simulations of homogeneous and isotropic Navier–Stokes turbulence.

Dissipation can be modeled in different ways, including joint friction or drag. Here, we consider a bending viscosity acting at the joints, which damps the relative angular motion of neighboring links.
To leading order, the dissipation operator $\mathcal D$ acts linearly on angular velocities (Appendix~\ref{app:operators}). It is tridiagonal, reflecting the locality of joint friction, and positive semidefinite, ensuring nonnegative energy dissipation. Moreover, $\mathcal D\mathbf 1=0$, because uniform rotation of all links involves no relative angular motion.

The resulting forced and dissipative dynamics reads
\begin{equation}
\label{eq:deterministic}
\dot{\boldsymbol\theta}=\boldsymbol\omega,\quad
\dot{\boldsymbol\omega}=\mathcal A(\boldsymbol\theta)^{-1}
\big[\boldsymbol b(\boldsymbol\theta,\boldsymbol\omega)
+\boldsymbol{\tau}-\eta\mathcal D\boldsymbol\omega\big],
\end{equation}
where $\eta$ is the viscosity parameter.
The dissipated power,
\begin{equation}
\eta\boldsymbol\omega^T\mathcal D\boldsymbol\omega
=\eta\sum_{k=2}^{N}\frac{(\omega_k-\omega_{k-1})^2}{\ell_k},
\label{eq:dissipated-power}
\end{equation}
weights the smallest links most strongly, just as Laplacian viscosity preferentially dissipates energy at the smallest scales of a turbulent flow.
The dissipative forces also preserve the total angular momentum, which therefore remains an invariant of the full dynamics \eqref{eq:deterministic}.

Unlike the Euler equations for an incompressible velocity field, our system has a canonical Hamiltonian structure, with positions and momenta as independent variables. Since the kinetic energy metric depends on position \eqref{eq:lagr2}, the dissipation operator and the kinetic energy cannot, in general, be diagonalized in the same fixed basis. In incompressible Navier–Stokes turbulence, by contrast, both kinetic energy and viscous dissipation are diagonal in Fourier space.
Nevertheless, turbulent behavior is expected to exhibit universal properties, independent of the specific mechanisms of forcing and dissipation. To investigate this robustness, we consider two forcing protocols.

The first is a regularized constant-power forcing,
\begin{equation}
\tau=\varepsilon_0
\frac{\omega_1-\omega_2}
{(\omega_1-\omega_2)^2+\omega_r^2},
\label{eq:forcing-constant-power}
\end{equation}
which injects power close to $\varepsilon_0$. The regularization frequency $\omega_r>0$ prevents a singularity when the relative angular velocity vanishes and is kept small. This protocol provides a convenient statistically stationary state (used throughout unless stated otherwise).

Because this forcing directly controls the injected power, however, it is not suitable for testing anomalous dissipation, namely whether the mean dissipation rate remains finite as $\eta\to0$.
For this purpose, we also consider a fixed-amplitude periodic torque
\begin{equation}
\tau=a_f\cos(\omega_f t),
\label{eq:forcing2}
\end{equation}
for which the injected power is determined by the response of the chain.
This forcing is also invariant under time reversal $t\mapsto-t$, a property we exploit when testing the irreversibility of the dynamics.

The dissipative joints can be promoted to thermal reservoirs by introducing fluctuations consistent with the fluctuation–dissipation relation, thereby extending the model to Langevin dynamics \cite{callenwelton1951,landau1987}.
The same operator $\mathcal D$ that governs friction also determines the thermal fluctuations. Writing $\mathcal D=\mathcal G^T\mathcal G$, the full stochastic dynamics reads
\begin{equation}
\mathcal A\,d\boldsymbol\omega=
\big(\boldsymbol b-\eta\mathcal D\boldsymbol\omega+\boldsymbol\tau\big)dt
+\sqrt{2\eta\Theta}\,\mathcal G^T d\boldsymbol W,
\label{eq:langevin}
\end{equation}
where $\boldsymbol W$ is a vector of independent Wiener processes and $\Theta$ sets the thermal energy.
In terms of the canonical momentum $\boldsymbol p=\mathcal A\boldsymbol\omega$, Eq.~\eqref{eq:langevin} has the standard Langevin form, with stochastic torque covariance
\begin{equation}
    \langle d\boldsymbol\Gamma\,d\boldsymbol\Gamma^T\rangle=2\eta\Theta
    \,\mathcal D\,dt .
    \label{eq:noise-covariance}
\end{equation}
Dissipation and fluctuations are thus controlled by the same operator, as in Landau–Lifshitz fluctuating hydrodynamics \cite{landau1987,foxuhlenbeck1970}. Once $\eta$ and $\Theta$ are fixed, the noise amplitude is determined by fluctuation–dissipation balance.
Since $\mathcal D$ gives greater weight to the tail of the chain, thermal torque fluctuations are strongest at the smallest scales, while their amplitude vanishes as $\sqrt{\eta}$ in the inviscid limit.
The deterministic dynamics is recovered by setting $\Theta=0$. Since $\mathcal D\mathbf 1=0$ implies $\mathcal G\mathbf 1=0$, the stochastic torques sum to zero, and the total angular momentum is conserved by the stochastic dynamics as well.

\subsection{Nonequilibrium energy cascade}
\label{sec:scaling-prediction}

In a statistically stationary state, the mean energy injection rate $\varepsilon$ equals the average of the dissipation rate \eqref{eq:dissipated-power}.
If the scales of the energy source and sink are sufficiently separated, an inertial range of pure energy transfer can emerge, characterized by a constant mean transfer rate fixed by the global balance:
\begin{equation}
\label{eq:inertial_range}
\langle\Pi_k\rangle\simeq\varepsilon.
\end{equation}
This is the analogue of the constant mean energy flux across wavenumbers in three-dimensional Navier–Stokes turbulence \cite{kolmogorov1941,frisch1995turbulence}.
The nonzero energy flux distinguishes this statistically stationary state from thermal equilibrium, where the mean energy transfer across scales vanishes.
In a nonequilibrium stationary cascade, we assume that energy transfer across link $k$ is controlled by motions at comparable scales, in the spirit of Kolmogorov’s scale-locality hypothesis \cite{kolmogorov1941,obukhov1941}.
Introducing the characteristic velocity $u_k=\ell_k|\omega_k|$ and the corresponding transfer time $\ell_k/u_k$, a constant energy flux requires
$\varepsilon\sim m_k u_k^3/\ell_k=m_k\ell_k^2|\omega_k|^3$.
Since $m_k\propto\ell_k$, this yields
\begin{equation}
|\omega_k|\sim\ell_k^{-1},\qquad
e_k\sim\ell_k,\qquad
|j_k|\sim\ell_k.
\label{eq:kolmogorov-scaling}
\end{equation}
The scaling of $e_k$ and $j_k$, which are nonlocal observables in angular coordinates, is discussed in more detail in Appendix~\ref{app:variance-scalings}.
The characteristic velocities are therefore independent of scale, $u_k\sim\ell_k^0$, while the turnover times $\ell_k/u_k\propto\ell_k$ decrease geometrically along the chain.

The same dimensional balance predicts the end of the cascade. The dissipated power at link $k$ scales as
$\eta\omega_k^2/\ell_k\sim\eta\ell_k^{-3}$.
Equating this to the energy flux gives the dissipative scale
\begin{equation}
\ell_\eta\propto\eta^{1/3},
\label{eq:dissipative-scale}
\end{equation}
so that, for the hierarchical ratio $\gamma=2$, the inertial range extends by $\log_2 10^{1/3}\simeq1.1$ links for each decade of decrease in viscosity.

In thermal equilibrium, by contrast, energy is shared equally among the
degrees of freedom. With $\mathcal A_{kk}\sim\ell_k^3$, equipartition gives
(Appendix~\ref{app:equipartition})
\begin{equation}
    \lvert\omega_k\rvert\sim\ell_k^{-3/2},\qquad
    e_k\sim\ell_k^{0},\qquad
    \lvert j_k\rvert\sim\ell_k^{1/2}.
    \label{eq:equilibrium-scaling}
\end{equation}
The two states differ qualitatively. In equilibrium every mass carries the
same energy, so the total energy grows without bound as links are added,
the mechanical analogue of the ultraviolet catastrophe of classical
equipartition. In the cascade, the energy per link decreases
geometrically toward small scales and the total energy is finite, while
the energy flowing through each link remains constant. The results of
Sec.~\ref{sec:results-cascade} show which of the two the driven chain
selects.

\subsection{The hierarchy as a control parameter}
\label{sec:design}

The choice of hierarchy provides a way to control the scaling properties of the cascade. In particular, lengths and masses need not decrease at the same rate. If they follow independent geometric ratios, so that $m_k\propto\ell_k^{\alpha}$, the same dimensional argument predicts velocity increments
\begin{equation}
    u_k\sim\ell_k^{h_c},\qquad h_c=\frac{1-\alpha}{3}.
\label{eq:hc}
\end{equation}
The derivation is given in Appendix~\ref{sec:onsager_holder}.

For uniform masses ($\alpha=0$), this relation yields $h_c=1/3$, recovering the exponent of Kolmogorov’s theory and Onsager’s conjecture~\cite{onsager1949}. In this case, the energy flux per unit mass scales as $u_k^3/\ell_k$, exactly as in a fluid. As $\alpha$ increases, the predicted cascade becomes progressively rougher. The chain studied here, with $\alpha=1$, lies at the critical value $h_c=0$, where velocity increments become scale-independent. For $\alpha>1$, the exponent is negative, $h_c<0$, allowing singular velocity increments while the total energy remains finite.

Thus, unlike in a fluid, where the regularity of the cascade is fixed by the equations of motion, in this mechanical system it can be tuned through the design of the hierarchy.

\section{Results}
\label{sec:results}

The results follow the three stages announced in the introduction. We
first establish that the deterministic chain carries a cascade and
characterize how the energy is transferred: locally, irreversibly and
intermittently. We then add the thermal noise that any physical
realization would carry, and finally ask what survives as the viscous
regularization is removed. Unless stated otherwise, we integrate
Eq.~\eqref{eq:deterministic} with the constant-power forcing
\eqref{eq:forcing-constant-power}; the fixed-amplitude forcing
\eqref{eq:forcing2} is used for the irreversibility and dissipative-anomaly
tests, and Eq.~\eqref{eq:langevin} for the stochastic dynamics. All
statistics refer to statistically stationary states. Simulation parameters
are listed in Appendix~\ref{app:params}.

\subsection{Deterministic dynamics}
\label{sec:results-deterministic}

\begin{figure*}
    \centering
    \includegraphics[width=\textwidth]{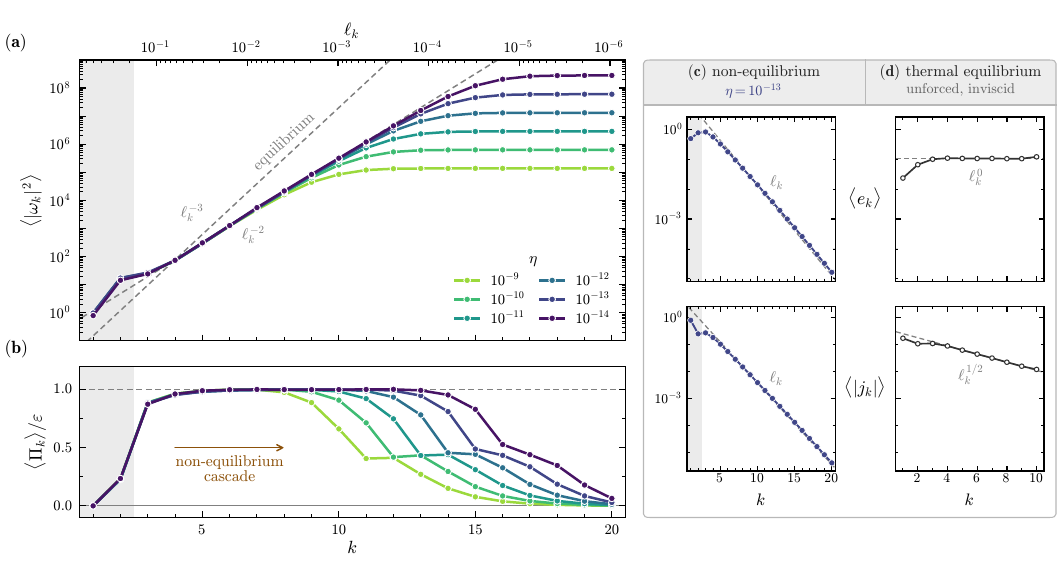}
    \caption{\textbf{A cascade far from equilibrium.}
    (a) Angular velocity spectrum for decreasing viscosity $\eta$, approaching the inertial-range scaling $\langle|\omega_k|^2\rangle\sim\ell_k^{-2}$; the steeper guide marks the equipartition scaling $\ell_k^{-3}$.
    (b) Mean energy flux normalized by the injection rate, exhibiting an increasingly extended constant-flux plateau.
    (c) Nonequilibrium mean kinetic energy $\langle e_k\rangle$ and mean absolute angular momentum $\langle|j_k|\rangle$ for $\eta=10^{-13}$, both following $\ell_k$.
    (d) The same observables in thermal equilibrium (unforced, inviscid chain), following $\ell_k^0$ and $\ell_k^{1/2}$.
    Shaded bands mark the forced links $k=1,2$.}
    \label{fig:noneq_eq}
\end{figure*}

\subsubsection{A cascade far from equilibrium}
\label{sec:results-cascade}

Does the driven chain cascade, or does it equipartition?
Figure~\ref{fig:noneq_eq} answers this by comparing the
driven-dissipative chain with the same chain in thermal equilibrium. As
$\eta$ decreases, the angular velocity spectrum in Fig.~\ref{fig:noneq_eq}(a) develops an extended range following the cascade
prediction
\begin{equation}
    \langle|\omega_k|^2\rangle\propto\ell_k^{-2}.
    \label{eq:result-omega2}
\end{equation}
The existence of a cascade is established
directly by the flux. As shown in Fig.~\ref{fig:noneq_eq}(b), after the first few links, which are dominated by
the forcing, the mean flux reaches a plateau
$\langle\Pi_k\rangle\simeq\varepsilon$. The
plateau widens as the viscosity decreases, its small-scale edge moving by
about one link per decade of $\eta$, from $k\simeq9$ at $\eta=10^{-9}$ to
$k\simeq14$ at $\eta=10^{-14}$, as predicted by
Eq.~\eqref{eq:dissipative-scale}. Beyond the dissipative scale the
spectrum flattens: the smallest links co-rotate rigidly, a motion that the
bending viscosity, which only damps relative rotation, cannot remove.
Movie~S1 shows the driven chain in this statistically stationary state
\cite{supplemental}, and an interactive version runs in the web
browser \cite{website}.
The cascade spectrum is much shallower than the equipartition
spectrum $\langle|\omega_k|^2\rangle\propto\ell_k^{-3}$ of
Eq.~\eqref{eq:equilibrium-scaling}, shown for comparison in
Fig.~\ref{fig:noneq_eq}(a): in equilibrium the small scales would rotate
much faster than they do in the cascade.

The contrast with equilibrium is sharpest in how the energy is
distributed among the links. Figures~\ref{fig:noneq_eq}(c) and
\ref{fig:noneq_eq}(d) contrast the
kinetic energy and angular momentum of each mass in the driven chain with
those of the unforced, inviscid chain after relaxing from energy injected in
its first two links. The relaxed chain equipartitions,
$\langle e_k\rangle\sim\ell_k^0$ and
$\langle|j_k|\rangle\sim\ell_k^{1/2}$, as in
Eq.~\eqref{eq:equilibrium-scaling}: left to itself, the energy injected at
the largest scales spreads evenly over all links. The driven state instead
follows Eq.~\eqref{eq:kolmogorov-scaling},
$\langle e_k\rangle\sim\langle|j_k|\rangle\sim\ell_k$, over the whole
chain, far from equipartition. This scaling alone, however, does not
identify a cascade. For the critical hierarchy $\alpha=1$ the cascade
velocities are scale independent, $u_k\sim\ell_k^0$, so $e_k$ and $|j_k|$
are simply proportional to the mass $m_k\propto\ell_k$, as they would be
for a chain rotating rigidly; this is why the same scaling persists in the
dissipative range, where the smallest links co-rotate. The energy and
angular momentum therefore rule out equilibrium but not rigid rotation.
Rigid rotation is ruled out by the angular velocity spectrum, which would
be flat instead of following $\ell_k^{-2}$, and by the flux, which would
vanish. The driven chain thus selects the cascade: a nonequilibrium
state, sustained by the flux, in which energy flows through the hierarchy
instead of being shared among its degrees of freedom.

\begin{figure*}
    \centering
    \includegraphics[width=\textwidth]{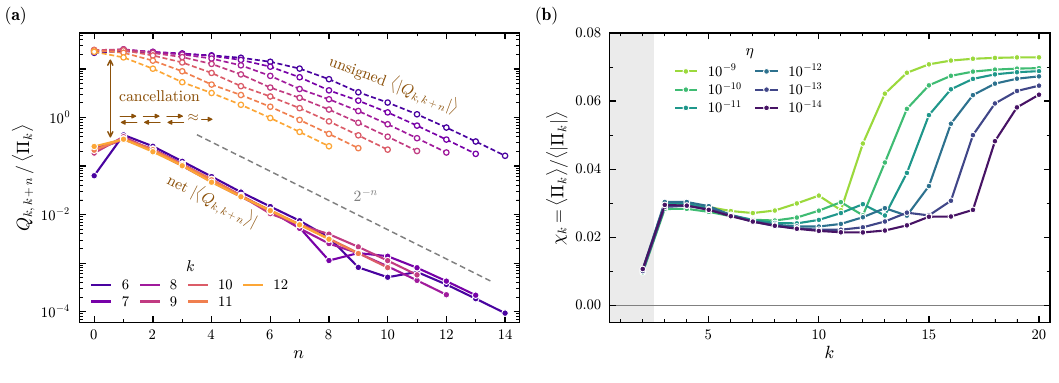}
    \caption{\textbf{Statistical locality and time irreversibility.}
    (a) Net, $|\langle Q_{k,k+n}\rangle|$ (filled), and unsigned, $\langle|Q_{k,k+n}|\rangle$ (open), contributions to the flux through cut $k$ from interactions across a scale separation $n$, normalized by the total $\langle\Pi_k\rangle$. The net contribution decays as $2^{-n}$, while remote interactions remain strong instantaneously but largely cancel in the mean.
    (b) Normalized flux bias $\chi_k=\langle\Pi_k\rangle/\langle|\Pi_k|\rangle$ for decreasing $\eta$, with the time-reversal-symmetric fixed-amplitude cosine forcing. In the inertial range the lowest-viscosity curves approach a common nonzero profile.}
    \label{fig:locality_irrev}
\end{figure*}

These results also settle a point of principle. The chain has no
prescribed coupling between neighboring scales: its equations are
nonlocal, and the cascade could have been short-circuited by direct
transfers from the forced links to the dissipative ones. The emergence of
an extended constant-flux range shows that the hierarchy organizes the
transfer by itself. How this happens is the subject of the next section.

\subsubsection{Local transfer from nonlocal coupling}
\label{sec:statistical_locality}

The cascade prediction assumes that energy is exchanged mainly between
comparable scales, yet the inertia couples every link to the whole chain.
To resolve this apparent contradiction,
we decompose the flux as
$\Pi_k=\sum_{j\geq k}Q_{k,j}$, where $Q_{k,j}$ is the contribution of mass
$j$ to the tension of rod $k$ (Appendix~\ref{app:flux}). We then measure,
for a scale separation $n$, the net contribution
$|\langle Q_{k,k+n}\rangle|$, which enters the mean cascade, and the
unsigned contribution $\langle|Q_{k,k+n}|\rangle$, which measures the
typical instantaneous exchange, both normalized by $\langle\Pi_k\rangle$
and shown in Fig.~\ref{fig:locality_irrev}(a).

The two measures behave very differently. Remote scales are strongly
coupled at every instant: the unsigned contribution decays slowly with $n$
and remains of the order of the mean flux over several links. Their net
contribution, however, decays as $2^{-n}$, and is already about two orders
of magnitude below the unsigned one at $n\simeq7$. Forward and backward exchanges with remote
scales cancel almost exactly in the mean, and a statistically local
cascade emerges from nonlocal dynamics. The distinction between dynamical
and statistical locality is central to hydrodynamic cascades
\cite{eyinkaluie2009}, where pressure plays the role of the tension here.
In the chain, the rate of decay follows from the mass hierarchy.
Writing
$Q_{k,k+n}=m_{k+n}\,\ddot{\mathbf r}_{k+n}\cdot\dot{\mathbf r}_k$,
Eq.~\eqref{eq:Qkj}, and assuming that the correlation
$\langle\ddot{\mathbf r}_{k+n}\cdot\dot{\mathbf r}_k\rangle$ saturates in
the inertial range, gives
$|\langle Q_{k,k+n}\rangle|\propto m_{k+n}\propto2^{-n}=(\ell_{k+n}/\ell_k)^{1}$.
Locality is thus inherited from the decreasing inertia of the small
scales, which can respond strongly to the large ones but cannot carry a
large net share of the flux.

Locality tells us where the energy goes; the same flux statistics also
tell us that it cannot come back. Irreversibility is a defining feature of turbulence, visible for instance
in the asymmetric statistics of energy gains and losses along fluid
trajectories \cite{xu2014flightcrash}. Under time reversal the velocities
change sign while the tensions do not, so $\Pi_k\mapsto-\Pi_k$, and a
time-symmetric stationary state would have $\langle\Pi_k\rangle=0$. A
nonzero mean flux therefore already implies irreversibility, but at finite
viscosity this is not surprising, since the viscous term breaks time
reversal explicitly. To isolate the dynamical origin of the asymmetry, we
use the periodic forcing of Eq.~\eqref{eq:forcing2}, which is itself
time-reversal symmetric, so that the equations break reversibility only
through the viscous term, and ask whether the asymmetry survives as
$\eta\to0$. We measure it through the flux bias
\begin{equation}
    \chi_k=\frac{\langle\Pi_k\rangle}{\langle|\Pi_k|\rangle},
\end{equation}
the net transfer relative to the magnitude of the instantaneous exchanges
in both directions. A finite mean flux could in principle be accompanied by
ever larger bidirectional exchanges, sending $\chi_k$ to zero. Instead,
$\chi_k$ converges in the inertial range to a common nonzero profile as
$\eta$ decreases, as Fig.~\ref{fig:locality_irrev}(b) shows, while the sharp rise
at larger $k$ moves to smaller scales with the dissipative cutoff. The
relative imbalance between forward and backward transfer persists as the
coefficient of the only explicitly irreversible term vanishes: the arrow of
time is set by the cascade, not by the regularization.

\begin{figure*}
    \centering
    \includegraphics[width=\textwidth]{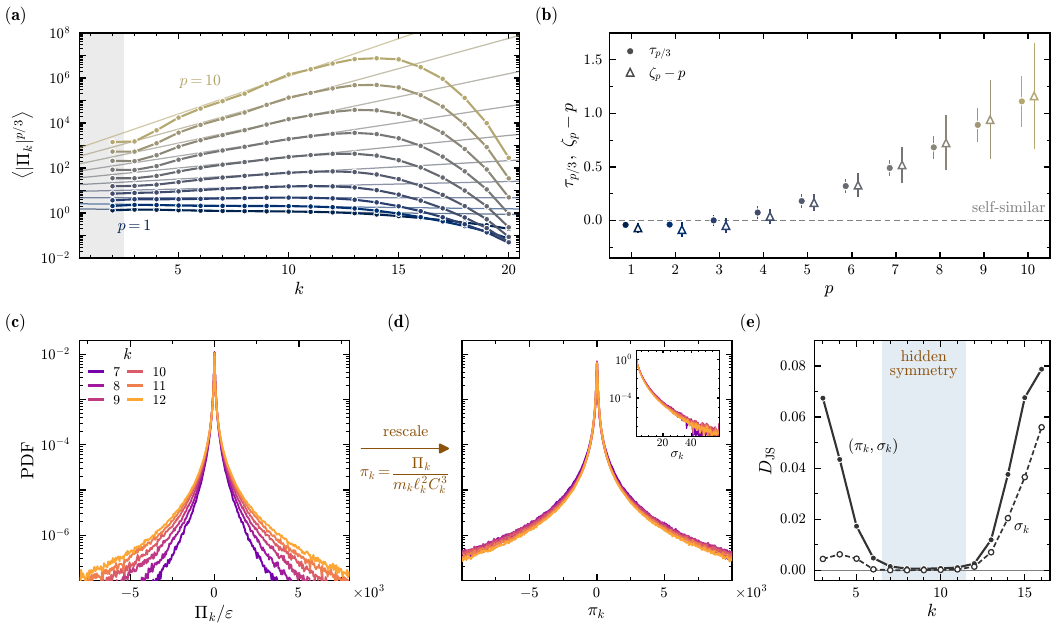}
    \caption{\textbf{Intermittency and hidden symmetry.} Results for $\eta=10^{-13}$.
    (a) Moments $\langle|\Pi_k|^{p/3}\rangle$ of the instantaneous energy flux for $p=1,\dots,10$, with fitted power laws (thin lines).
    (b) Anomalous corrections obtained independently from flux moments, $\tau_{p/3}$, and from angular velocity moments, $\zeta_p-p$; self-similar scaling corresponds to zero.
    (c) Probability distributions of the energy flux at different cuts $k$, showing the absence of collapse.
    (d) Collapse of the normalized flux $\pi_k=\Pi_k/(m_k\ell_k^2C_k^3)$, computed using the velocity amplitudes $C_k$ predicted by the hidden-symmetry theory. Inset: collapse of the amplitude multipliers $\sigma_k=C_{k+1}/C_k$.
    (e) Jensen--Shannon divergence between the distributions at adjacent scales, for the joint distribution of $(\pi_k,\sigma_k)$ and for $\sigma_k$ alone. The shaded band marks the range over which the hidden symmetry is realized.}
    \label{fig:intermittency_hs}
\end{figure*}

\subsubsection{Intermittency and hidden symmetry}
\label{sec:results-intermittency}

So far we have characterized the mean transfer. The mean flux fixes the
second-order statistics, but not the fluctuations, which in turbulence
are where self-similarity breaks down. We define scaling exponents
through
\begin{equation}
    \langle|\omega_k|^p\rangle\propto\ell_k^{-\zeta_p},
    \qquad
    \langle|\Pi_k|^{p/3}\rangle\propto\ell_k^{-\tau_{p/3}} .
    \label{eq:exponents}
\end{equation}
The cascade prediction gives $\zeta_p=p$, so that $\zeta_p-p$ measures the
departure from self-similarity. Since $u_k=\ell_k\omega_k$, positive
corrections $\zeta_p-p>0$ correspond to the usual negative corrections for
velocity increments, $\langle|u_k|^p\rangle\sim\ell_k^{p-\zeta_p}$. The
flux moments, taken to the power $p/3$ because the flux is cubic in the
velocity, provide an independent determination. If the transfer is local,
$|\Pi_k|\sim m_k\ell_k^2|\omega_k|^3$, and the two sets of exponents are
related by
\begin{equation}
    \tau_{p/3}=\zeta_p-p,
    \label{eq:refined-relation}
\end{equation}
the form of refined similarity proposed by Kraichnan
\cite{kraichnan1974}, in which the fluctuations of the local energy
transfer rate, rather than those of the dissipation rate
\cite{kolmogorov1962}, control the anomalous velocity statistics
\cite{eyink1995}.

Figure~\ref{fig:intermittency_hs}(a) shows the flux moments at
$\eta=10^{-13}$, which follow clean power laws over the inertial range.
In Fig.~\ref{fig:intermittency_hs}(b), both anomalous corrections vanish near $p=3$, as required by the constant
mean flux, and grow systematically at higher orders, where the statistics are dominated
by increasingly rare and violent events.
The two determinations agree over
all orders within uncertainties, estimated conservatively from both
statistical and systematic sources \cite{dewit2024extreme}. This agreement
is not trivial. Anomalous exponents measured from $\omega_k$ alone could
reflect the choice of angular coordinates or the nonlocal inertia of the
chain. Their independent recovery from the exact mechanical flux shows that
intermittency is carried by the fluctuations of the interscale energy
transfer itself. The chain thus displays not only a mean cascade but an
intermittent one.

Intermittency breaks the dimensional scaling symmetry: as Fig.~\ref{fig:intermittency_hs}(c) shows, the flux
distributions at different scales do not collapse. This does not mean that scale
invariance is lost altogether. A symmetry of the equations need not be
a symmetry of the attractor, as the broken dimensional scaling shows, but
the attractor may still realize a scale symmetry in other variables.
Following the hidden-symmetry framework of
Refs.~\cite{mailybaev2021hidden,mailybaev2022hidden,mailybaev2022spatiotemporal},
we normalize each configuration by the instantaneous amplitude of the
larger scales,
\begin{equation}
    C_k=\Big(\sum_{j<k}\omega_j^2\Big)^{1/2},
\end{equation}
and measure time with the corresponding intrinsic clock (Appendix~\ref{app:hidden}). The rescaled
dynamics is exactly invariant under the shift $k\mapsto k+1$, combined
with a renormalization of amplitudes and time. This symmetry is hidden, because it acts on
the projected and not on the original variables, and the projection is
nonlinear. Whether the turbulent state realizes it is a dynamical
question. If it does, the amplitude
multipliers and normalized fluxes,
\begin{equation}
    \sigma_k=\frac{C_{k+1}}{C_k},\qquad
    \pi_k=\frac{\Pi_k}{m_k\ell_k^2C_k^3},
    \label{eq:multipliers}
\end{equation}
must have scale-independent statistics.

Indeed they do. Rescaling the flux by the hidden-symmetry amplitudes collapses the
distributions of Fig.~\ref{fig:intermittency_hs}(c) onto a single curve, shown in Fig.~\ref{fig:intermittency_hs}(d), and the multiplier distributions
collapse likewise, as shown in the inset. The Jensen--Shannon divergence between the
distributions at adjacent cuts, plotted in Fig.~\ref{fig:intermittency_hs}(e), quantifies the collapse.
It is smallest over
the inertial range, which identifies the range over which the joint statistics
of $(\pi_k,\sigma_k)$, and therefore the hidden symmetry, are realized. Intermittency thus does not destroy scale
invariance; it replaces ordinary self-similarity with a symmetry that is
revealed only by a nonlinear change of variables. This has a predictive
consequence. Because the multipliers are statistically self-similar, the
structure functions take a multiplicative form whose scaling is governed by
the leading eigenvalue of a positive transfer operator built from the
multiplier statistics \cite{mailybaev2022hidden,mailybaev2022spatiotemporal}.
The anomalous exponents are then determined by correlations among multipliers.

\begin{figure}
    \centering
    \includegraphics[width=\columnwidth]{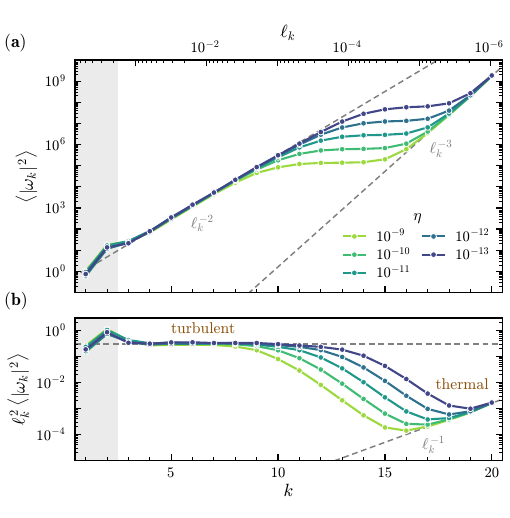}
    \caption{\textbf{Turbulent-to-thermal crossover.} (a) Stationary angular velocity spectra with fluctuation--dissipation-consistent noise, for decreasing $\eta$. The turbulent scaling $\langle|\omega_k|^2\rangle\sim\ell_k^{-2}$ is retained at large scales, while the smallest scales approach the thermal equipartition scaling $\ell_k^{-3}$.
    (b) The same spectra compensated by $\ell_k^{2}$: a turbulent plateau, a viscous range that moves to smaller scales as $\eta$ decreases, and a common, $\eta$-independent thermal range.}
    \label{fig:crossover}
\end{figure}

\subsection{Stochastic dynamics}
\label{sec:results-thermal}

Any physical realization of the chain is subject to thermal
fluctuations, whose amplitude is here fixed by the dissipation. They are
the natural microscopic perturbation for spontaneous stochasticity, although
any vanishingly small perturbation would serve. Before
asking how they propagate, we check what they do to the stationary state. Figure~\ref{fig:crossover} shows
the stationary spectra of the stochastic dynamics,
Eq.~\eqref{eq:langevin}. The large scales retain the cascade scaling
$\ell_k^{-2}$, while the smallest scales relax to the equipartition
spectrum $\ell_k^{-3}$ of Eq.~\eqref{eq:equilibrium-scaling}. The
compensated spectra in Fig.~\ref{fig:crossover}(b) separate three ranges: a
turbulent plateau; a viscous range, which moves to smaller scales as $\eta$
decreases; and a thermal range, whose level is set by $\Theta$ alone and is
independent of $\eta$.

The thermal tail confirms that the noise is the equilibrium counterpart of
the bending viscosity, as fluctuation--dissipation balance requires
\cite{callenwelton1951,landau1987}. The same principle underlies
discretizations of fluctuating hydrodynamics \cite{bell2007,donev2010,delong2013},
and an analogous thermal range has been predicted in the far-dissipation
range of fluid turbulence \cite{bandak2022}. The rigid-rotation mode,
which is neither damped nor forced, constrains the dynamics to a fixed
angular momentum, but this only adds a rank-one correction to the Gibbs
covariance and does not affect the ultraviolet scaling
(Appendix~\ref{app:equipartition}).
The energy and canonical momentum,
which depend nonlocally on all larger links, retain their turbulent
scalings over the scales studied (Appendix~\ref{app:variance-scalings}).
Thermal noise is therefore not an external source of randomness. It is
set by the dissipation, confined to the smallest scales, leaves the
cascade intact, and vanishes as $\sqrt\eta$. If it nevertheless affects the
large scales, it must be because the turbulent dynamics amplifies it.

\subsection{Singular limits}
\label{sec:results-singular}

We can now ask the central question: what survives as the regularization
is removed? Two properties of turbulence are singular in this sense. The
dissipative anomaly concerns the mean energy budget, and spontaneous
stochasticity concerns the predictability of individual trajectories.
The chain studied here exhibits both.

\begin{figure*}
    \centering
    \includegraphics[width=\textwidth]{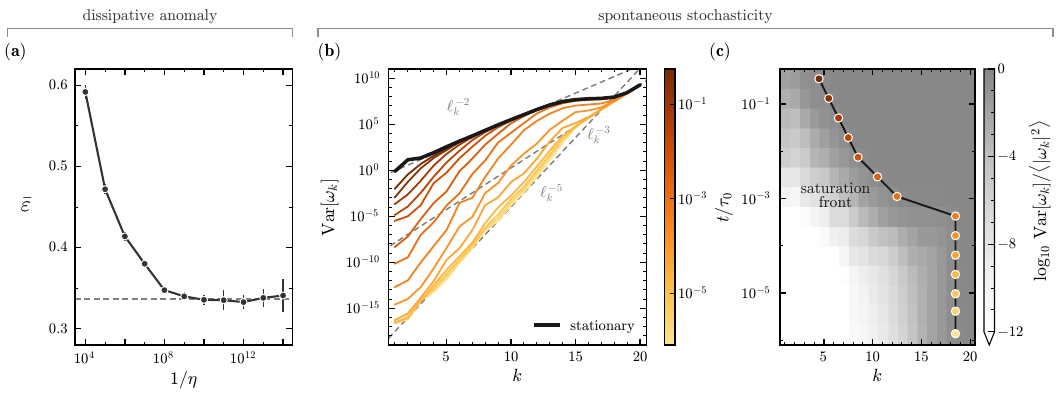}
    \caption{\textbf{Singular dynamics in the limit $\eta\to0$.}
    (a) Dissipative anomaly: mean stationary energy dissipation rate $\bar\varepsilon$ as a function of inverse viscosity for the fixed-amplitude forcing. As the viscosity is reduced, $\bar\varepsilon$ approaches a finite nonzero plateau (dashed line).
    (b,c) Spontaneous stochasticity: upscale propagation of microscopic uncertainty for $\eta=10^{-13}$, deep in the range where $\bar\varepsilon$ has saturated. (b) Ensemble variance of the angular velocities at successive times $t/\tau_0$, approaching the stationary spectrum (thick line); dashed lines indicate the short-time ($\ell_k^{-5}$), thermal ($\ell_k^{-3}$) and turbulent ($\ell_k^{-2}$) reference scalings. (c) The same variances normalized by their stationary values, in the $(k,t)$ plane. The line marks the saturation front, where the variance first reaches half of its stationary value; its markers carry the colors of the corresponding curves in (b).
    }
    \label{fig:singular}
\end{figure*}

\subsubsection{Dissipative anomaly}
\label{sec:results-anomalous-dissipation}

With the constant-power forcing of Eq.~\eqref{eq:forcing-constant-power}, the mean
dissipation is imposed by the forcing and cannot test the anomaly. We
therefore use the fixed-amplitude forcing of Eq.~\eqref{eq:forcing2}, hold
its parameters fixed, and let the stationary state determine the injected
power. A strong dissipative anomaly corresponds to
\begin{equation}
    \lim_{\eta\to0}\bar\varepsilon(\eta)=\varepsilon_*>0 .
    \label{eq:strong-anomaly}
\end{equation}
Figure~\ref{fig:singular}(a) shows the mean dissipation rate
$\bar\varepsilon$ over ten decades of $\eta$. At moderate viscosities it
varies appreciably, but as $\eta$ decreases it saturates at a finite
plateau, $\bar\varepsilon\simeq0.34$, with no systematic decay over the
smallest viscosities explored.

The mechanism is the one at work in fluids. By the energy balance, a finite
limit requires
$\langle\boldsymbol\omega^T\mathcal D\boldsymbol\omega\rangle\sim\eta^{-1}$:
increasingly violent relative rotations of ever smaller links compensate
the vanishing viscous prefactor. This is consistent with the cascade
picture, in which the dissipative scale $\ell_\eta\propto\eta^{1/3}$ moves
to smaller scales at a fixed flux. The distinction between the strong and
weak forms of the anomaly is subtle in three-dimensional turbulence, where
it remains unsettled \cite{zeroth_law}. Within the accessible range, the
pendulum chain realizes the strong form, in a mechanically explicit system
where the dissipation is a well-defined and measurable power.

\subsubsection{Spontaneous stochasticity}
\label{sec:results-spst}

The second singular feature concerns predictability: if two copies of the
chain start from exactly the same state, how long do they remain
alike when the only difference between them is thermal noise? To isolate
the effect of the noise from the chaotic variability of the attractor, we
use a two-level ensemble (Appendix~\ref{app:protocol}). We draw $64$
independent states from the stationary deterministic dynamics, and from
each we evolve $128$ solutions of Eq.~\eqref{eq:langevin} that start from
exactly the same configuration and differ only in their noise realizations.
The variance $\operatorname{Var}[\omega_k](t)$ across noise realizations,
averaged over the initial states, then measures the uncertainty generated
by the noise alone. Time is measured in units of the large-scale period
$\tau_0=2\pi/\langle\omega_1^2\rangle^{1/2}$, and $\eta=10^{-13}$.
Spontaneous stochasticity corresponds to a variance that remains finite at
fixed $k$ and $t$ in the joint limit $\eta\to0$, $N\to\infty$
\cite{mailybaev2016,eyinkdrivas2015,mailybaev2016nonlinear,eyinkbandak2020},
even though the noise amplitude vanishes in that limit.

Figures~\ref{fig:singular}(b) and \ref{fig:singular}(c) show that uncertainty is born at the smallest scales and propagates up the hierarchy,
against the direction of the energy cascade. The variance
profiles reflect three regimes, whose reference scalings are derived in
Appendix~\ref{app:variance-scalings}. At the earliest times, the variance
grows linearly in time as the direct response to the noise, with a steep
profile ($\ell_k^{-5}$ empirically). The smallest scales then relax to the
thermal covariance, $\ell_k^{-3}$. Finally, as the memory of the common
initial condition is lost, the variance saturates at the stationary
turbulent level, $\ell_k^{-2}$. The crossover to saturation moves
systematically toward larger scales. The saturation front of
Fig.~\ref{fig:singular}(c), where the variance first reaches half its
stationary value, climbs the hierarchy from the dissipative scale
toward the forced links, each step to a larger link taking longer than
the previous one. Measured in units of the local period
$\tau_k=2\pi/\langle\omega_k^2\rangle^{1/2}$, however, the arrival times
are comparable at all scales: the front reaches each link after a time comparable to its own period, advancing on the local turnover time. By
$t\simeq0.3\,\tau_0$, the uncertainty
born at the dissipative scale has climbed to the large scales of the
chain.

This structure distinguishes the mechanism from ordinary chaos. In a system
with a single characteristic time, a perturbation of size $\delta$ grows
as $\delta e^{t/\tau}$ and needs a time $\tau\log(1/\delta)$, which
diverges as $\delta\to0$, to become macroscopic. In a hierarchy, the
uncertainty instead climbs one scale at a time, each step taking a time of
the order of the local turnover time, $\ell_k/u_k\propto\ell_k$. These
times form a geometric series dominated by the largest scales, so the
total time needed to reach the top of the hierarchy remains finite however
small the scale at which the uncertainty is born. This is Lorenz's
mechanism for a finite predictability horizon \cite{lorenz1969}, developed
in scale-dependent predictability theories
\cite{leithkraichnan1972,aurell1996} and observed in Navier--Stokes and
shell-model turbulence \cite{boffetta2017,budanur2022,bandak2024,freitas2026stochasticity}.
It is also what turns vanishing noise into a finite effect in the limit:
as $\eta\to0$ the noise weakens, but the hierarchy extends to smaller and
faster scales that amplify it faster.

\begin{figure*}
    \centering
    \includegraphics[width=\textwidth]{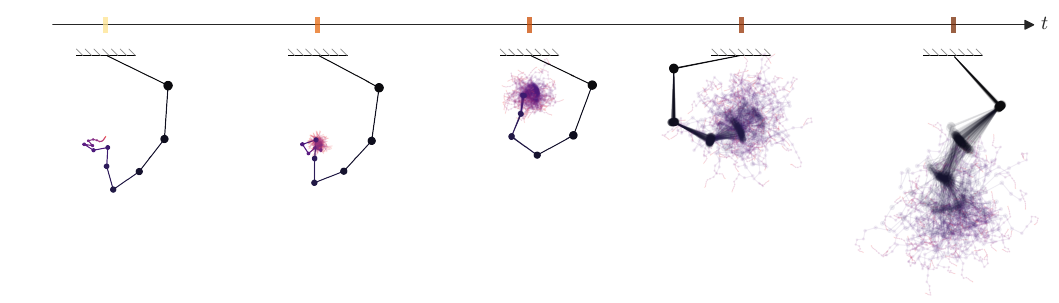}
    \caption{\textbf{Visualization of upscale uncertainty propagation.} Snapshots of 128 stochastic realizations evolving from an identical initial condition under independent noise realizations. The initially localized uncertainty progressively spreads from the smallest to the largest links, providing a physical-space visualization of the inverse error propagation quantified in Fig.~\ref{fig:singular}. The ticks on the time axis mark the snapshot times, colored as the corresponding times $t/\tau_0$ in Fig.~\ref{fig:singular}(b,c). For visual clarity, the link-length ratio differs from that used in the simulations. The full time evolution is shown in Movie~S2 \cite{supplemental}.}
    \label{fig:sptsto_viz}
\end{figure*}

Figure~\ref{fig:sptsto_viz} shows the same process in physical space.
Chains that start from an identical configuration and differ only by
their thermal noise remain indistinguishable at the large scales while
their tips spread; the spread then climbs the chain until the realizations
differ at every scale. Movie~S2 follows this process continuously in time
\cite{supplemental}.

\section{Discussion}
\label{sec:discussion}

\subsection{A minimal recipe for turbulence}

A chain of rigid pendula with geometrically decreasing lengths and masses
reproduces every hallmark of turbulence we have tested: a constant-flux
cascade far from equilibrium, statistically local and irreversible energy
transfer, intermittency with a hidden scale symmetry, a finite dissipation
rate as the viscosity vanishes, and the amplification of thermal noise from
the smallest to the largest scales in a finite time. None of these
properties is put in by hand. The coupling follows from rigid constraints
and inertia; it is nonlocal and conserves energy and angular momentum
without any designed triadic structure.

The answer to the question posed in the introduction can therefore be
stated as a recipe with three ingredients. The first is a
\emph{hierarchy of inertia}, which provides the separation of length and
time scales. The second is a \emph{nonlinear coupling} across the
hierarchy, without which the degrees of freedom would not exchange energy;
here it is purely geometric, inherited from the rigid constraints, and
present at every amplitude. The third is a \emph{separation between
injection and dissipation}, large-scale forcing and small-scale damping,
which opens an inertial range in between. The last two are plainly
necessary for any cascade. The first is not, since in fluids the hierarchy
emerges from the dynamics, but our results show that imposing it is enough:
the three together are sufficient. Incompressibility, spatial structure and vortex dynamics shape
turbulence in fluids, but they are not required for its universal singular
features.
Several of our results make this concrete. Locality of the transfer, which
in fluids is often attributed to the structure of the nonlinearity, here
emerges from the decreasing inertia of the small scales, through the
cancellation of instantaneous nonlocal exchanges. Irreversibility is
selected by the cascade itself and persists when the explicitly
irreversible term vanishes. And intermittency is organized by the same
hidden symmetry found in shell models and in Navier--Stokes turbulence
\cite{mailybaev2022spatiotemporal,mailybaevthalabard2022,calascibetta2025hidden},
which suggests that hidden self-similarity is a generic property of
hierarchical cascades rather than of particular equations.

The hierarchy also bears on finite-time singularities. A finite chain
cannot blow up, but with infinitely many links and no viscosity the
turnover times form a convergent series (Sec.~\ref{sec:results-spst}), so
energy can reach arbitrarily small scales in a finite time. This is the
mechanism of blow-up in dyadic models \cite{katz2005dyadic,cheskidov2008dyadic}
and in Tao's averaged Navier--Stokes equations \cite{tao2016averaged}, and
it underlies constructions of singular solutions of the forced Euler and
Boussinesq equations from flows stacked at ever smaller scales
\cite{cordoba2023euler,cordoba2025boussinesq}. The chain realizes it with a purely geometric nonlinearity and without vortex stretching.

\subsection{Turbulence by design}

Because the hierarchy is the essential ingredient, it is also the natural
control parameter. The mass and length ratios set the dimensional exponent
of the cascade, Eq.~\eqref{eq:hc}, from Kolmogorov's $1/3$ for uniform
masses to singular increments for heavy hierarchies. The forcing,
dissipation and noise act through explicit operators that can be chosen
independently, and the dissipative scale and the thermal range are set by
$\eta$ and $\Theta$ through simple balances. This shifts turbulence from a
property one observes to a property one can design. Natural questions
follow: how the intermittency corrections and the range of the hidden
symmetry depend on $\alpha$, and which features of turbulence can be
switched off by breaking the hierarchy in a controlled way.
Another concerns the ratio $\gamma$ itself, which plays the role of the
inter-shell spacing: in shell models, reducing the spacing enhances
intermittency through the phase synchronization of consecutive
triads~\cite{Manfredini2026}, and the chain allows one to test whether the
same mechanism operates in a mechanical hierarchy.

Design also means realization. Since only the inertia matrix enters the
conservative dynamics, the same Lagrangian can be built from very
different components. We give an electromechanical realization in which
capacitances play the role of masses and magnetic flux linkages that of
lengths (Appendix~\ref{app:experimental}); there, the hierarchy is set by
component values rather than by physical sizes, which relaxes the
mechanical difficulty of building a chain over many decades of scale. The
ideal correspondence requires rotor inertia, coil inductance and losses to
remain negligible.
In either realization, every quantity
entering our analysis, including positions, velocities and constraint
forces, is in principle directly measurable. This opens the way to
laboratory studies of the singular phenomena of turbulence, in particular
of spontaneous stochasticity, whose direct observation in fluids is
notoriously challenging.

\subsection{Beyond the chain}

The recipe suggests a family of designed turbulent systems, obtained
by changing one ingredient at a time. Start with the hierarchy. In the chain each scale has a single degree of
freedom, so energy can only pass from one link to the next, whereas in a
fluid the number of degrees of freedom grows toward small scales and energy
can flow through many parallel channels. Branching hierarchies, in which
each rigid body carries two or more smaller copies of itself, reproduce this
growth and turn the chain into a tree. Their inertia matrix remains dense
along each genealogical line and admits the same exact linear-cost algebra,
so the effect of branching on locality, intermittency and the hidden
symmetry can be studied directly, much as in multi-branch shell models
\cite{tuteri2026fixedpoints,tuteri2026multibranch2d,tuteri2026intermittency}.
Movie~S3 shows a first simulation of such a tree \cite{supplemental}.

One can also ask how much of the hierarchy is needed. In fluids the
separation of time scales is emergent, following from the spatial cascade
through the turnover times; in the chain it follows from the imposed
lengths and masses. The logic can be reversed by imposing a hierarchy in
time alone: identical copies of the same system, each running on its own
clock, with clocks scaled geometrically from one level to the next. Such systems
contain no spatial structure at all, and would test whether a hierarchy of
time scales is by itself sufficient for a cascade and its singular
properties.

The nonlinearity can be varied as well. The rigid chain has no
restoring force: its angles follow a free geodesic flow, with no linear
regime. Torsional springs at the joints, or gravity, add a potential energy
and with it linear waves, whose frequency relative to the local turnover
rate becomes a control parameter. Stiff springs confine the links to small
oscillations, where weakly interacting waves should exchange energy through
resonances, as in wave turbulence \cite{zakharov1992,nazarenko2011}; soft
springs leave the links free to rotate and should recover the strong
turbulence studied here. A hierarchical elastic chain would let one follow
the transition between the two regimes, possibly with both coexisting at
different scales. Its uniform counterpart, a pendulum chain with gravity and
torsional coupling, is the classic mechanical analogue of the sine-Gordon
equation \cite{scott1969}, so the hierarchical version would also bring
coherent nonlinear excitations and turbulent cascades into the same system.

Springs that act along their length, connecting the masses as in the
Fermi--Pasta--Ulam--Tsingou chain \cite{fermi1955}, offer another route, in
which the nonlinearity need not be built into the springs at all: in two or three
dimensions, the length of a spring depends nonlinearly on the displacements
of its ends, so nested networks of Hookean masses and springs, or nested
rigid bodies coupled by elastic joints, are mechanical metamaterials whose
nonlinearity is again purely geometric.

Finally, the recipe takes for granted that the couplings are
reciprocal, as Newton's third law
demands for passive mechanical systems. Active or feedback-controlled
realizations can break this reciprocity, as in odd elastic and
non-reciprocal active materials
\cite{scheibner2020odd,fruchart2021nonreciprocal}, and might reveal odd
phenomena with no counterpart in ordinary turbulence~\cite{dewit2024pattern}.

\subsection{Outlook}

For the chain itself, several questions remain. The strict spontaneous-stochastic limit requires
the convergence of the large-scale statistics as $\eta\to0$ with
increasing $N$, and the chain, with its exact $O(N)$ dynamics, is well
suited to this test. Its relation to finite-time Lyapunov exponents, and
to renormalization-group descriptions of singular cascades
\cite{eyinkbandak2020,mailybaev2026rgsabra,mailybaevmoriconi2026}, remains
to be established. The dissipative anomaly calls for a rigorous
counterpart of Onsager's criterion for the exact, nonlocal mechanical
flux, which Appendix~\ref{sec:onsager_holder} formulates at the
dimensional level. Finally, it would be interesting to simulate a time-reversible
version of the chain, as done for reversible shell models of turbulence
\cite{biferale2018equivalence}: whether it sustains a stationary cascade
with the same statistics would clarify the origin of the arrow of time in
turbulence.
More broadly, the chain provides a minimal, physical and tunable setting in
which the ingredients of turbulence can be added or removed one at a time.
Turbulence, long studied as a property of fluids, emerges here as a
property of hierarchies, and one that can be designed.

\begin{acknowledgments}
The authors acknowledge useful discussions with Alexandros Alexakis, Luca Biferale, Sergio Chibbaro, Xander de Wit, Alexei Mailybaev and Vincenzo Vitelli.
FT acknowledges financial support from the CNRS through the MITI interdisciplinary initiatives, under its exploratory research program.
This work was granted access to the HPC resources of GENCI-TGCC under Project No. A0190506421.
AF acknowledges support from the European Union’s HORIZON MSCA Doctoral Networks programme under Grant Agreement No. 101072344, project AQTIVATE (Advanced computing, QuanTum algorIthms and data-driVen Approaches for science, Technology and Engineering), the European Research Council (ERC) under the European Union’s Horizon 2020 research and innovation programme Smart-TURB (Grant Agreement No. 882340).
\end{acknowledgments}

\section*{Data availability}
The numerical simulation code as well as the data used to produce the figures will be made available upon publication. An interactive version of the simulation is archived at Ref.~\cite{website}.

\appendix
\section{Energy flux and its scale decomposition}
\label{app:flux}

In Cartesian coordinates, the equations of motion of the conservative system \eqref{eq:conservative-compact} read
\begin{equation}
    m_k\ddot{\mathbf r}_k
    =\lambda_k(\mathbf r_k-\mathbf r_{k-1})-\lambda_{k+1}(\mathbf r_{k+1}-\mathbf r_k),
    \label{eq:cartesian-eom}
\end{equation}
where $\lambda_k$ are the Lagrange multipliers enforcing the rigidity constraints, with $\lambda_{N+1}=0$.
Summing these equations from $k$ to $N$ yields
\begin{equation}
    \lambda_k(\mathbf r_k-\mathbf r_{k-1})=\sum_{j=k}^{N}m_j\ddot{\mathbf r}_j.
    \label{eq:tension}
\end{equation}
Thus, the tension force in link $k$ is determined by the accelerations of all the masses it supports.
Taking the scalar product of Eq.~\eqref{eq:cartesian-eom} with $\dot{\mathbf r}_k$ and using the differentiated inextensibility constraint
$(\mathbf r_{k+1}-\mathbf r_k)\cdot(\dot{\mathbf r}_{k+1}-\dot{\mathbf r}_k)=0$
gives the local energy budget \eqref{eq:local-budget}.

Equation~\eqref{eq:tension} also provides a decomposition of the energy flux into contributions from different interaction ranges. Substituting it into the definition of $\Pi_k$ gives
\begin{equation}
    \Pi_k=\sum_{j=k}^{N}Q_{k,j},\qquad
    Q_{k,j}=m_j\ddot{\mathbf r}_j\cdot\dot{\mathbf r}_k.
    \label{eq:Qkj}
\end{equation}
Each term $Q_{k,j}$ represents the contribution associated with mass $j$ beyond link $k$.
This decomposition is used in Fig.~\ref{fig:locality_irrev}(a) to assess the locality of the energy cascade.

\section{Dissipation and noise operators}
\label{app:operators}

For notational convenience, we introduce the discrete bending rate,
defined as the rate of change of discrete curvature:
\begin{equation}
    \delta\omega_k
    =\frac{\omega_k-\omega_{k-1}}{\ell_k},
    \qquad k=2,\ldots,N.
\end{equation}
We assume that bending friction is linear in the discrete bending rate
to leading order. Each link in the bulk receives two contributions
from its adjacent joints, so that, given the friction parameter $\eta$, the corresponding torque has components
\begin{equation}
    d_k=-\eta(\delta\omega_k-\delta\omega_{k+1}),
    \qquad k=2,\ldots,N-1,
    \label{eq:dissipation-interior}
\end{equation}
while for the end points
$d_1=\eta\,\delta\omega_2$ and $d_N=-\eta\,\delta\omega_N$.
These expressions are linear in the angular velocities and can then be expressed in terms of a
symmetric tridiagonal matrix
\begin{equation}
\mathcal D=
\begin{pmatrix}
g_1 & f_1 & & & \\
f_1 & g_2 & f_2 & & \\
& \ddots & \ddots & \ddots & \\
& & f_{N-2} & g_{N-1} & f_{N-1} \\
& & & f_{N-1} & g_N
\end{pmatrix},
\label{eq:D-matrix}
\end{equation}
where
\begin{equation}
    g_j=\frac{1-\delta_{j1}}{\ell_j}
    +\frac{1-\delta_{jN}}{\ell_{j+1}},
    \qquad
    f_j=-\frac{1}{\ell_{j+1}}.
\end{equation}
With these definitions, the dissipative torque vector is given by $\mathbf d=-\eta\mathcal D\boldsymbol\omega$.
The matrix $\mathcal D$ is positive semidefinite and has $\mathbf 1$, corresponding to rigid rotation, in its kernel, so that uniform rotations are not dissipated.

To construct a Langevin dynamics consistent with the fluctuation--dissipation relation, we need a square root of the dissipation matrix. A factorization $\mathcal D=\mathcal G^T\mathcal G$ is
\begin{equation}
\mathcal G_{ak}
=\frac{\delta_{a,k}}{\sqrt{\ell_a}}(1-\delta_{k,1})
-\frac{\delta_{a,k+1}}{\sqrt{\ell_a}},
\label{eq:G}
\end{equation}
where $a,k=1,\ldots,N$.
With this construction, the stochastic dynamics is driven by the bath noise torque $\mathcal G^T d\boldsymbol W$, as in Eq.~\eqref{eq:langevin}, so that the covariance operator is indeed the friction matrix $\mathcal{D}$.

\section{Hidden-symmetry projection}
\label{app:hidden}

A symmetry of the equations of motion need not be realized by the statistical properties of the turbulent chaotic attractor. In particular, intermittency
observed in our numerical simulations
(Sec.~\ref{sec:results-intermittency}) shows that the scaling symmetries
\eqref{eq:time_scaling} and \eqref{eq:scale_shift}, although exact at the
level of the equations, do not translate into statistical self-similarity.

Following Refs.~\cite{mailybaev2021hidden,mailybaev2022spatiotemporal}, we seek a different symmetry, one that is realized by the turbulent dynamics despite the breaking of the original scaling symmetries. This hidden symmetry emerges in a reduced space obtained by normalizing the dynamics with state-dependent amplitudes and intrinsic clocks. In these variables, statistical self-similarity can be recovered even though it is absent in the original description.

To define this symmetry, we focus on the bulk, where boundary effects can be neglected, and consider then an infinite chain, $k\in\mathbb Z$, with square-summable angular velocities.
By recentering the chain, we may choose the reference site to be $k=0$ without loss of generality and define the amplitude
\begin{equation}
A(\boldsymbol{\omega})=\sqrt{\sum_{k<0}\omega_k^2}.
\end{equation}
The key idea is to normalize the angular velocities to unit amplitude and rescale time using the amplitude itself:
\begin{equation}
\theta_k\mapsto\theta_k,
\quad
\omega_k\mapsto\Omega_k=\frac{\omega_k}{A(\boldsymbol{\omega})},
\quad
d\tau=A(\boldsymbol{\omega})\,dt.
\end{equation}
The resulting projected dynamics is
\begin{align}
    \mathcal A(\boldsymbol\theta)\frac{d\boldsymbol\Omega}{d\tau}
    &=\boldsymbol b(\boldsymbol\theta,\boldsymbol\Omega)
    -\alpha\,\mathcal A(\boldsymbol\theta)\boldsymbol\Omega,
    \qquad
    \frac{d\boldsymbol\theta}{d\tau}=\boldsymbol\Omega,
    \label{eq:proj}
    \\
    \alpha&=\sum_{k<0}\Omega_k\big[\mathcal A(\boldsymbol\theta)^{-1}
    \boldsymbol b(\boldsymbol\theta,\boldsymbol\Omega)\big]_k.
\end{align}
The projected system indeed admits a scaling symmetry, as can be verified by direct substitution:
\begin{equation}
\theta_k\mapsto\theta_{k+1},\qquad
\Omega_k\mapsto\frac{\Omega_{k+1}}{C_1},\qquad
d\tau\mapsto C_1\,d\tau,
\label{eq:sym-implementation}
\end{equation}
where $C_1^2=\sum_{k<1}\Omega_k^2$.
This transformation is the hidden form (hidden in the projected space) of the scale-shift symmetry $k\mapsto k+1$ of Eq.~\eqref{eq:scale_shift}.

Iterating the shift, we obtain the discrete family of scale symmetries, characterized by the amplitudes
\begin{equation}
C_m=\sqrt{\sum_{k<m}\Omega_k^2}.
\end{equation}
The statistical realization of the symmetry \eqref{eq:sym-implementation} is tested through the multipliers, amplitude ratios, and normalized fluxes defined in Eq.~\eqref{eq:multipliers}, whose statistics are accumulated using the intrinsic clock $\tau$.

As a technical detail, for a finite chain, a set of dummy links with negative indices is introduced to realize the projection onto $A=1$, so that the normalized quantities coincide with those computed directly from $\omega_k$.

As a final remark, statistical self-similarity also allows intermittency to be interpreted as a multiplicative process with multipliers exhibiting scale-invariant statistics, even though they are generally correlated~\cite{mailybaev2022hidden,mailybaev2022spatiotemporal,calascibetta2025hidden}.

\section{Ensemble protocol}
\label{app:protocol}

We construct the ensemble by sampling $N_{\rm IC}=64$ statistically independent initial states
\begin{equation}
    \mathbf z_0^{(\alpha)}=(\boldsymbol\theta_0^{(\alpha)},\boldsymbol\omega_0^{(\alpha)}),\qquad\alpha=1,\ldots,N_{\rm IC}
\end{equation}
from the stationary deterministic dynamics. Starting from each of these states, we evolve
$N_\xi=128$ solutions of Eq.~\eqref{eq:langevin}, each driven by an independent
Wiener process. For a given observable $X_k$, we first compute the conditional variance across noise realizations,
\begin{equation}
    V_{X,k}^{(\alpha)}(t)=\frac{1}{N_\xi}\sum_{r=1}^{N_\xi}
    \Big[X_k^{(\alpha,r)}(t)-\overline{X}_k^{(\alpha)}(t)\Big]^2,
\end{equation}
where $\overline{X}_k^{(\alpha)}(t)$ denotes the corresponding conditional mean.
We then average these variances over the sampled initial states,
$V_{X,k}(t)=N_{\rm IC}^{-1}\sum_\alpha V_{X,k}^{(\alpha)}(t)$.
This procedure isolates the uncertainty generated by the noise from the variability associated with sampling different states of the attractor.

A spontaneously stochastic limit corresponds to a nonvanishing conditional variance as the noise amplitude vanishes and the number of links tends to infinity:
\begin{equation}
\liminf_{\eta\to0,\,N\to\infty}V_{\omega,k}(t)>0,
\qquad k,\ t,\ \Theta\ \text{fixed}.
\label{eq:spst-limit}
\end{equation}
This can occur when the zero-viscosity limit is singular: rather than converging to a single solution of the inviscid conservative equations, the dynamics converges to a probability distribution over solutions, selected by the limiting procedure itself. Figure~\ref{fig:singular}(b,c) illustrates this behavior for parameter values that numerically appear to realize the spontaneously stochastic limit for the angular velocities, consistently with the independent evidence provided by anomalous dissipation.

\section{Realizations}
\label{app:experimental}

\begin{figure*}
    \centering
    \includegraphics[width=0.98\textwidth]{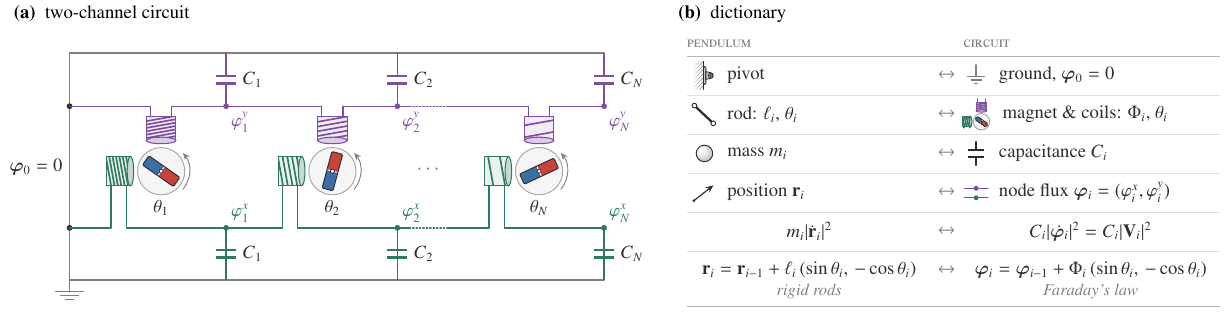}
    \caption{\textbf{Electromechanical realization.}
    (a)~Two-channel circuit, corresponding to the two components in the plane, $x$ and $y$.
    Each link of the chain consists of a permanent magnet that can rotate freely (angular degree of freedom $\theta_i$), with its orientation determining the magnetic flux through two orthogonal coils.
    The coil with a vertical axis links a flux proportional to $-\cos\theta_i$ and belongs to the $y$ channel (upper line), while the coil with a horizontal axis links a flux proportional to $\sin\theta_i$ and belongs to the $x$ channel (lower line). In each channel, the coil at stage $i$ connects node $i-1$ to node $i$, and each node is connected to ground through a capacitor $C_i$. The chain begins at the grounded node $\bm\varphi_0=0$, which plays the role of the pivot, and terminates at $C_N$. The components are scaled along the chain but are identical in the two channels. The hierarchy is determined entirely by the capacitances and the number of coil turns, which sets the flux-linkage amplitude $\Phi_i$.
    (b)~One-to-one correspondence between the pendulum chain and the circuit. The node flux $\boldsymbol{\varphi}_i=(\varphi^x_i,\varphi^y_i)$ plays the role of the position $\mathbf r_i$ of mass $i$, while the capacitive energy $C_i|\dot{\boldsymbol{\varphi}}_i|^2$ corresponds to the kinetic energy $m_i|\dot{\mathbf r}_i|^2$. Faraday’s law across stage $i$ gives
    $\boldsymbol{\varphi}_i=\boldsymbol{\varphi}_{i-1}+\Phi_i(\sin\theta_i,-\cos\theta_i)$.
    This reproduces the rigid-rod constraint under the identification $\ell_i\to\Phi_i$, so that the circuit Lagrangian maps exactly onto that of the hierarchical pendulum chain, with $m_i\to C_i$.}
    \label{fig:electromechanical-realization}
\end{figure*}

The conservative dynamics of the gravity-free pendulum chain is determined by the configuration-dependent inertia matrix $\mathcal A$ in Eq.~\eqref{eq:metric}. Its geometrical structure can therefore be reproduced by systems involving different physical interactions. Here, we describe a hybrid electromechanical realization in which capacitive energy plays the role of mechanical kinetic energy, while rotating magnets coupled to coils impose the analogues of rigid-link constraints. In the ideal limit, this construction reproduces the pendulum Lagrangian~\eqref{eq:lagr2}, with the hierarchy determined by the capacitances and magnetic flux-linkage amplitudes.

The underlying analogy identifies masses with capacitances, linear
velocities with voltages, and forces with currents. A complete dictionary is illustrated in
Fig.~\ref{fig:electromechanical-realization}(b). To represent the
two Cartesian components of the planar chain, we introduce two
electrical channels, $x$ and $y$, as shown in
Fig.~\ref{fig:electromechanical-realization}(a). At each stage $i$, a
magnet rotating about a fixed axis couples to two orthogonal coils,
one in each channel. Within each channel, the coils are connected
in series, and the node following stage $i$ is connected to the common
reference through a capacitor. The two capacitors at that stage have
the same capacitance $C_i$.

Let $\varphi_i^\alpha$ denote the node flux in channel
$\alpha=x,y$, defined by
$V_i^\alpha=\dot\varphi_i^\alpha$, where $V_i^\alpha$ is the
corresponding capacitor voltage. The reference fixes
$\varphi_0^x=\varphi_0^y=0$.
Assuming that the two orthogonal coils
at stage $j$ have the same flux-linkage amplitude $\Phi_j$,
Faraday induction and the series connections give
\begin{equation}
    \varphi_i^x=\sum_{j=1}^{i}\Phi_j\sin\theta_j,
    \qquad
    \varphi_i^y=-\sum_{j=1}^{i}\Phi_j\cos\theta_j.
    \label{eq:electrical-flux}
\end{equation}
Here $\theta_j$ is measured from the negative $y$ direction in the
rotor plane, following the convention of
Eq.~\eqref{eq:coordinates}. Equation~\eqref{eq:electrical-flux}
has exactly the same form as the pendulum kinematics, with
$\Phi_j$ replacing $\ell_j$.

Here we neglect rotor inertia, coil self-inductance, and losses.
Apart from losses, which are excluded from a conservative cascade
by definition, these corrections can be included in the description,
provided they scale appropriately. What is fundamental, regardless
of the physical mechanisms involved, is the scaling symmetry
of the equations of motion.

The Lagrangian is given by the capacitive contribution
\begin{equation}
    \mathcal L_{\mathrm{el}}
    =
    \frac12\sum_{i=1}^{N}C_i
    \left[
        (\dot\varphi_i^x)^2+(\dot\varphi_i^y)^2
    \right]
    =
    \frac12\dot{\boldsymbol{\theta}}\cdot
    \mathcal A^{\mathrm{el}}(\boldsymbol\theta)\dot{\boldsymbol{\theta}},
    \label{eq:electrical-lagrangian}
\end{equation}
where differentiation of Eq.~\eqref{eq:electrical-flux} gives
\begin{equation}
    \mathcal A^{\mathrm{el}}_{kj}(\boldsymbol\theta)
    =
    \Phi_k\Phi_j\cos(\theta_j-\theta_k)
    \sum_{i=\max(k,j)}^{N}C_i.
    \label{eq:electrical-metric}
\end{equation}
This is the same inertial structure as
Eqs.~\eqref{eq:lagr2} and~\eqref{eq:metric} under the
identifications $\ell_i\to\Phi_i$ and $m_i\to C_i$.
In particular, the hierarchy of Eq.~\eqref{eq:hierarchy}
corresponds to
\begin{equation}
    \Phi_{i+1}=\Phi_i/\gamma,\qquad
    C_{i+1}=C_i/\gamma,
    \qquad \gamma>1.
\end{equation}
Remarkably, gravity produces no torque about the rotation axes
of the magnets, so gravitational effects are absent in this
realization, without approximation.

\section{Numerical parameters}
\label{app:params}

Unless stated otherwise, all simulations use $N=20$ links and the geometric
hierarchy of Eq.~\eqref{eq:hierarchy}, with $\gamma=2$ and the normalization
$m_1=1$, $\ell_1=1$.

For the deterministic simulations, we use an adaptive time step based on the
two characteristic time scales associated with angular velocity and angular
acceleration,
\begin{equation}
\Delta t
=
\beta
\min\left\{
    \left\|\boldsymbol{\omega}\right\|_\infty^{-1},
    \left(
        \left\|\dot{\boldsymbol{\omega}}_{\rm free}\right\|_\infty
        + 1
    \right)^{-1/2}
\right\},
\label{eq:adaptive-dt}
\end{equation}
where $\beta=0.1$, $\dot{\boldsymbol{\omega}}_{\rm free}$ denotes
the angular acceleration due to the free dynamics, and the addition
of $1$ prevents divergence of the second time scale
when the acceleration becomes small.
The viscous contribution is excluded from Eq.~\eqref{eq:adaptive-dt}
because it is treated implicitly.

For the ensemble calculations of
Sec.~\ref{sec:results-spst}, all trajectories are instead advanced with the
fixed time step
$\Delta t_{\rm sto}=5\times10^{-7}$,
so that all ensemble members are evaluated at identical physical times.

For the anomalous dissipation test of
Sec.~\ref{sec:results-anomalous-dissipation}, the near-constant-power forcing
is replaced by Eq.~\eqref{eq:forcing2}. The forcing parameters are held fixed
at $a_f=0.1$ and $\omega_f=0.01$ for all viscosities.

\begin{table}[t]
\centering
\caption{Main numerical parameters.}
\label{tab:numerical-parameters}
\begin{tabular}{lll}
\toprule
Parameter & Value & Description \\
\midrule
$N$ & $20$ & number of links \\
$\varepsilon_0$ & $0.1$ & standard energy injection rate \\
$\omega_r$ & $0.1$ & forcing regularization \\
$\eta$ & $10^{-4}$--$10^{-14}$ & bending viscosity \\
$\beta$ & $0.1$ & adaptive time-step safety factor \\
$\Theta$ & $10^{-8}$ & fluctuation energy \\
\bottomrule
\end{tabular}
\end{table}

\section{Linear-complexity matrix operations}
\label{app:numerics}

The equations of motion \eqref{eq:deterministic} and \eqref{eq:langevin} require repeated solutions of linear systems involving
the configuration-dependent inertia matrix $\mathcal A(\boldsymbol{\theta})$.
Although $\mathcal A$ is dense, its hierarchical structure permits exact
factorizations and linear solves with computational cost $O(N)$.

\subsection{Semiseparable structure of the inertia matrix}
\label{app:semiseparable}

For $k\geq j$, the mass tensor satisfies
$M_{kj}=M_{kk}\equiv M_k$. Introducing
\begin{equation}
    c_j=\ell_j\cos\theta_j,
    \qquad
    s_j=\ell_j\sin\theta_j,
\end{equation}
and
\begin{equation}
    \widetilde c_k=M_k\ell_k\cos\theta_k,
    \qquad
    \widetilde s_k=M_k\ell_k\sin\theta_k,
\end{equation}
the lower triangular part of the matrix in
Eq.~\eqref{eq:metric} can be written as
\begin{equation}
    \mathcal A_{kj}
    =
    \widetilde c_k c_j
    +
    \widetilde s_k s_j,
    \qquad k\geq j.
    \label{eq:semiseparable-A}
\end{equation}
The upper triangular part follows by symmetry.
Thus only four vectors are required to represent the dense matrix, and its
storage cost is $O(N)$.

Since $\mathcal A$ is the kinetic-energy metric of the constrained system,
it is symmetric positive definite. We seek an exact factorization
\begin{equation}
    \mathcal A=LHL^T,
    \label{eq:LDL}
\end{equation}
where $L$ is unit lower triangular and
\begin{equation}
    H=\operatorname{diag}(h_1,\ldots,h_N).
\end{equation}
The semiseparable structure is inherited by $L$:
\begin{equation}
    L_{kj}
    =
    \widetilde c_k\alpha_j
    +
    \widetilde s_k\beta_j,
    \qquad k>j,
    \label{eq:L-semiseparable}
\end{equation}
with $L_{jj}=1$.
We build $L$ by induction. Suppose that the first $j-1$ columns of $L$ and the corresponding diagonal
entries of $H$ are known. Define
\begin{equation}
    U_j=\sum_{i<j}h_i\alpha_i^2,
    \quad
    V_j=\sum_{i<j}h_i\alpha_i\beta_i,
    \quad
    W_j=\sum_{i<j}h_i\beta_i^2.
    \label{eq:UVW}
\end{equation}
The $j$th pivot is
\begin{equation}
    h_j
    =
    \widetilde c_jc_j+\widetilde s_js_j
    -
    \left(
    \widetilde c_j^{\,2}U_j
    +2\widetilde c_j\widetilde s_jV_j
    +\widetilde s_j^{\,2}W_j
    \right).
    \label{eq:pivot}
\end{equation}
The off-diagonal coefficients follow from
\begin{align}
    c_j
    &=
    h_j\alpha_j
    +\widetilde c_jU_j
    +\widetilde s_jV_j,
    \\
    s_j
    &=
    h_j\beta_j
    +\widetilde s_jW_j
    +\widetilde c_jV_j.
    \label{eq:alphabeta}
\end{align}
Finally,
\begin{align}
    U_{j+1}&=U_j+h_j\alpha_j^2,
    \label{eq:UVW-update1}
    \\
    V_{j+1}&=V_j+h_j\alpha_j\beta_j,
    \label{eq:UVW-update2}
    \\
    W_{j+1}&=W_j+h_j\beta_j^2,
    \label{eq:UVW-update3}
\end{align}
starting from
\begin{equation}
    U_1=V_1=W_1=0.
\end{equation}
Each step requires a fixed number of scalar operations, so the complete
factorization has cost $O(N)$.

Consequently, the system
\begin{equation}
    \mathcal A\mathbf x=\mathbf b
\end{equation}
is solved without explicitly forming either $\mathcal A$ or
$\mathcal A^{-1}$.
Writing
\begin{equation}
    L\mathbf y=\mathbf b,
    \qquad
    H\mathbf z=\mathbf y,
    \qquad
    L^T\mathbf x=\mathbf z,
\end{equation}
the forward substitution takes the form
\begin{equation}
    y_k
    =
    b_k-\widetilde c_kP_k-\widetilde s_kQ_k,
    \label{eq:forward-A}
\end{equation}
where
\begin{equation}
    P_k=\sum_{j<k}\alpha_jy_j,
    \qquad
    Q_k=\sum_{j<k}\beta_jy_j.
\end{equation}
The diagonal solve is simply
\begin{equation}
    z_k=\frac{y_k}{h_k}.
\end{equation}
Backward substitution gives
\begin{equation}
    x_k
    =
    z_k-\alpha_kR_k-\beta_kS_k,
    \label{eq:backward-A}
\end{equation}
with
\begin{equation}
    R_k=\sum_{j>k}\widetilde c_jx_j,
    \qquad
    S_k=\sum_{j>k}\widetilde s_jx_j.
\end{equation}
The cumulative quantities in the forward and backward substitutions are
updated recursively, so each solve also requires $O(N)$ operations.
An analogous Gaussian-elimination strategy applies to linear
combinations of the inertia and dissipation matrices discussed
below, which arise in the implicit treatment of the dissipative term.

\subsection{Viscously shifted systems}
\label{app:shifted}

Both the implicit deterministic integrator and the stochastic thermostat
require solving systems of the form
\begin{equation}
    \mathcal B\mathbf x=\mathbf b,
    \qquad
    \mathcal B
    =
    \mathcal A+\rho\mathcal D,
    \label{eq:shifted-system}
\end{equation}
where $\rho$ is a scalar and $\mathcal D$ is the tridiagonal bending
operator defined in Sec.~\ref{sec:forcing-dissipation}.
The addition of $\rho\mathcal D$ preserves linear complexity, through a slightly more complicated Gaussian-elimination algorithm.

For the lower triangular factor we use the ansatz
\begin{equation}
    L_{kj}
    =
    \widetilde c_k\alpha_j
    +
    \widetilde s_k\beta_j
    +
    \delta_{k,j+1}\gamma_j,
    \qquad k>j.
    \label{eq:L-shifted}
\end{equation}
Using the notation of Eq.~\eqref{eq:UVW}, the corresponding recursions are
\begin{align}
    h_j
    &=
    \widetilde c_jc_j+\widetilde s_js_j+\rho g_j
    -
    \left(
    \widetilde c_j^{\,2}U_j
    +2\widetilde c_j\widetilde s_jV_j
    +\widetilde s_j^{\,2}W_j
    \right)
    \notag\\
    &-
    h_{j-1}\gamma_{j-1}
    \left[
        \gamma_{j-1}
        +2(
        \widetilde c_j\alpha_{j-1}
        +\widetilde s_j\beta_{j-1})
    \right],
    \label{eq:shifted-pivot}
    \\
    c_j
    &=
    h_j\alpha_j
    +\widetilde c_jU_j
    +\widetilde s_jV_j
    +h_{j-1}\gamma_{j-1}\alpha_{j-1},
    \\
    s_j
    &=
    h_j\beta_j
    +\widetilde s_jW_j
    +\widetilde c_jV_j
    +h_{j-1}\gamma_{j-1}\beta_{j-1},
    \\
    \rho f_j&=h_j\gamma_j,
    \label{eq:shifted-recursions}
\end{align}
with
\begin{equation}
    h_0=\alpha_0=\beta_0=\gamma_0=0.
\end{equation}
We recall that $g_j$ and $f_j$ are, respectively, the diagonal
and off-diagonal entries of $\mathcal D$ as shown in Eq.~\eqref{eq:D-matrix}.
The quantities $U_j,V_j,W_j$ are updated as in
Eqs.~\eqref{eq:UVW-update1},~\eqref{eq:UVW-update2},~\eqref{eq:UVW-update3}.
The additional first-subdiagonal contribution only modifies the forward and
backward substitutions locally:
\begin{equation}
    y_k
    =
    b_k
    -
    \widetilde c_kP_k
    -
    \widetilde s_kQ_k
    -
    \gamma_{k-1}y_{k-1},
    \label{eq:forward-shifted}
\end{equation}
and
\begin{equation}
    x_k
    =
    z_k
    -
    \alpha_kR_k
    -
    \beta_kS_k
    -
    \gamma_kx_{k+1},
    \label{eq:backward-shifted}
\end{equation}
with $\gamma_N=0$.
Consequently, factorization and solution of
Eq.~\eqref{eq:shifted-system} both remain $O(N)$.

\section{Deterministic time integration}
\label{app:deterministic-integrator}

For the driven-dissipative dynamics we introduce the variables
\begin{equation}
    \mathbf z
    =
    \begin{bmatrix}
        \boldsymbol{\theta}\\
        \boldsymbol{\omega}
    \end{bmatrix},
\end{equation}
and split the evolution separating the dissipative part, which will be solved implicitly:
\begin{equation}
    \dot{\mathbf z}
    =
    \mathbf F(t,\mathbf z)
    +
    \mathbf R(\mathbf z),
    \label{eq:IMEX-split}
\end{equation}
with
\begin{equation}
    \mathbf F(t,\mathbf z)
    =
    \begin{bmatrix}
        \boldsymbol{\omega}\\[2mm]
        \mathcal A(\boldsymbol{\theta})^{-1}
        [\boldsymbol b(\boldsymbol{\theta},\boldsymbol{\omega})
        +\boldsymbol{\tau}]
    \end{bmatrix},
\end{equation}
and
\begin{equation}
    \mathbf R(\mathbf z)
    =
    \begin{bmatrix}
        \mathbf 0\\[2mm]
        -\eta
        \mathcal A(\boldsymbol{\theta})^{-1}
        \mathcal D\boldsymbol{\omega}
    \end{bmatrix}.
\end{equation}
The nonlinear inertial dynamics and forcing are treated explicitly, whereas
the viscous term is treated implicitly.

As time integrators, we use the third-order IMEX-SSP3(4,3,3) Runge--Kutta scheme of
Ref.~\cite{pareschi2005imex}.
Let
\begin{equation}
    \mathbf Z_i
    =
    \begin{bmatrix}
        \boldsymbol{\Theta}_i\\
        \boldsymbol{\Omega}_i
    \end{bmatrix}
\end{equation}
denote the stage variables.
Because the implicit operator, i.e. the dissipative part, has no component in the angle equation,
\begin{equation}
    \boldsymbol{\Theta}_i
    =
    \boldsymbol{\theta}^{(n)}
    +
    \Delta t
    \sum_{j<i}
    a^E_{ij}\boldsymbol{\Omega}_j.
    \label{eq:IMEX-theta}
\end{equation}
Define the already known contribution at a given stage
\begin{equation}
    \mathbf q_i
    =
    \boldsymbol{\omega}^{(n)}
    +
    \Delta t
    \sum_{j<i}a^E_{ij}
    \mathbf f_j
    +
    \Delta t
    \sum_{j<i}a^I_{ij}
    \mathbf r_j,
    \label{eq:IMEX-q}
\end{equation}
where
\begin{align}
    \mathbf f_j
    &=
    \mathcal A(\boldsymbol{\Theta}_j)^{-1}
    \left[
        \boldsymbol b(\boldsymbol{\Theta}_j,\boldsymbol{\Omega}_j)
        +\boldsymbol{\tau}_j
    \right],
    \\
    \mathbf r_j
    &=
    -\eta
    \mathcal A(\boldsymbol{\Theta}_j)^{-1}
    \mathcal D\boldsymbol{\Omega}_j,
\end{align}
and $\boldsymbol{\tau}_j$ denotes the forcing evaluated at the
corresponding stage time and state.
The implicit stage is then obtained from
\begin{equation}
    \left[
        \mathcal A(\boldsymbol{\Theta}_i)
        +
        \Delta t\,a^I_{ii}\eta\mathcal D
    \right]
    \boldsymbol{\Omega}_i
    =
    \mathcal A(\boldsymbol{\Theta}_i)\mathbf q_i.
    \label{eq:IMEX-stage}
\end{equation}
This is precisely a shifted system of the form
Eq.~\eqref{eq:shifted-system} and is therefore solved in $O(N)$ operations.
After all stages are computed,
\begin{align}
    \boldsymbol{\theta}^{(n+1)}
    &=
    \boldsymbol{\theta}^{(n)}
    +
    \Delta t
    \sum_i b_i^E\boldsymbol{\Omega}_i,
    \\
    \boldsymbol{\omega}^{(n+1)}
    &=
    \boldsymbol{\omega}^{(n)}
    +
    \Delta t
    \sum_i b_i^E\mathbf f_i
    +
    \Delta t
    \sum_i b_i^I\mathbf r_i.
    \label{eq:IMEX-update}
\end{align}
Thus the complete deterministic time step retains linear complexity in the
number of links.

\section{Stochastic time integration}
\label{app:stochastic-integrator}

The stochastic dynamics in Eq.~\eqref{eq:langevin} contains two
qualitatively different components: the nonlinear driven mechanical
dynamics and a linear fluctuation--dissipation thermostat.
We integrate them using a symmetric Strang decomposition,
\begin{equation}
\Phi_{\Delta t}
=
\Phi_{\mathrm{th}}^{\Delta t/2}
\circ
\Phi_{\mathrm{mech}}^{\Delta t}
\circ
\Phi_{\mathrm{th}}^{\Delta t/2},
\label{eq:strang}
\end{equation}
where $\Phi_{\mathrm{mech}}$ advances
\begin{align}
d\boldsymbol{\theta}
&=
\boldsymbol{\omega}\,dt,
\\
\mathcal A(\boldsymbol{\theta})
d\boldsymbol{\omega}
&=
[\boldsymbol b+\boldsymbol{\tau}]\,dt,
\label{eq:mechanical-step}
\end{align}
and $\Phi_{\mathrm{th}}$ advances the viscous stochastic dynamics with
$\boldsymbol{\theta}$ frozen. The mechanical substep is evaluated with
a fourth-order explicit Runge--Kutta method.

For the thermostat substep we use the exact finite-time transition of the
frozen-configuration Ornstein--Uhlenbeck process. The required matrix
functions are evaluated using AAA high-precision rational approximations while retaining
$O(N)$ complexity.

\subsection{Ornstein--Uhlenbeck propagator}
\label{app:exact-OU}

During each thermostat substep the configuration
$\boldsymbol{\theta}$, and hence the inertia matrix
$\mathcal A(\boldsymbol{\theta})$, is held fixed. The thermostat equation is
therefore
\begin{equation}
\mathcal A\,d\boldsymbol{\omega}
=
-\eta\mathcal D\boldsymbol{\omega}\,dt
+
\sqrt{2\eta\Theta}\,
\mathcal G^T d\boldsymbol W.
\label{eq:OU}
\end{equation}
Defining
\begin{equation}
B=\eta\mathcal A^{-1}\mathcal D,
\qquad
C=\Theta\mathcal A^{-1},
\label{eq:BC}
\end{equation}
the exact transition over an interval $h$ is
\begin{equation}
\boldsymbol{\omega}(t+h)
=
e^{-hB}\boldsymbol{\omega}(t)
+
\boldsymbol{\zeta},
\qquad
\boldsymbol{\zeta}
\sim
\mathcal N(\mathbf 0,\Sigma_h),
\label{eq:OU-solution}
\end{equation}
with covariance
\begin{equation}
\Sigma_h
=
2\eta\Theta
\int_0^h
e^{-sB}
\mathcal A^{-1}\mathcal D\mathcal A^{-1}
e^{-sB^T}\,ds.
\label{eq:OU-cov-int}
\end{equation}
Since
\begin{equation}
BC=CB^T
=
\eta\Theta\,
\mathcal A^{-1}\mathcal D\mathcal A^{-1},
\end{equation}
the covariance can be written as
\begin{equation}
\Sigma_h
=
C-e^{-hB}Ce^{-hB^T}.
\label{eq:OU-cov}
\end{equation}

Using the factorization
\begin{equation}
\mathcal A=LHL^T
\end{equation}
from Appendix~\ref{app:semiseparable}, we introduce
\begin{equation}
\mathbf y=H^{1/2}L^T\boldsymbol{\omega},
\end{equation}
and the symmetric positive-semidefinite matrix
\begin{equation}
B_s
=
\eta
H^{-1/2}
L^{-1}\mathcal D L^{-T}
H^{-1/2}.
\label{eq:Bs}
\end{equation}
The thermostat update then becomes
\begin{equation}
\mathbf y(t+h)
=
e^{-hB_s}\mathbf y(t)
+
\Theta^{1/2}
\left(I-e^{-2hB_s}\right)^{1/2}
\boldsymbol{\xi},
\label{eq:OU-y}
\end{equation}
with $\boldsymbol{\xi}\sim\mathcal N(\mathbf 0,I)$.
Thus, the thermostat can be evolved simply by sampling from a Gaussian distribution; only the matrix-function actions need to be evaluated numerically.

\subsection{Linear-cost evaluation with AAA}
\label{app:AAA}

The update in Eq.~\eqref{eq:OU-y} requires the functional calculus
of
\begin{equation}
f(z)=e^{-z},
\qquad
g(z)=\sqrt{1-e^{-2z}}
\label{eq:fg}
\end{equation}
for $hB_s$. Rather than forming these dense matrix functions explicitly, we
use rational approximations constructed with the AAA algorithm
\cite{nakatsukasa2018aaa},
\begin{align}
f(z)
&\simeq
c_\infty
+
\sum_{r=1}^{m_f}
\frac{a_r}{z-p_r},
\\
g(z)
&\simeq
d_\infty
+
\sum_{r=1}^{m_g}
\frac{b_r}{z-q_r}.
\label{eq:AAA}
\end{align}
Each resolvent corresponds then to a system of the form
Eq.~\eqref{eq:shifted-system}.
Using the Gaussian-elimination method developed above, the complete
thermostat update retains $O(N)$ complexity for fixed approximation accuracy.

The AAA approximations are constructed on
$z\in[10^{-20},10^{10}]$ with a uniform tolerance of $10^{-10}$ in the
infinity norm. This results in $m_f=12$ poles for $f$ and $m_g=46$ poles for
$g$. The larger order required for $g$ reflects its branch-point singularity
at the origin, whereas $f$ is analytic.

\section{Equilibrium and thermalization}
\label{app:equipartition}

Energy equipartition at equilibrium can be achieved either through relaxation of the conservative dynamics or, at small scales, through the thermostat coupling.

\subsection{Equilibrium scaling and numerical comparison}

To study the global equilibrium, we first consider relaxation under the conservative dynamics, in the absence of forcing, viscosity, and stochastic noise.
The equilibrium state is constrained by the total energy $E$ and angular momentum $J$, both fixed by the initial conditions. For example, one may consider $J=0$ and $e_k=0$ for $k>2$, so that energy is initially concentrated in the first two masses.
The system is then allowed to evolve freely toward equilibrium. Dimensional arguments can then be used to predict the scaling of the equilibrium spectra.

For the geometrical hierarchy,
\begin{equation}
M_{kk}\sim m_k\sim\ell_k,
\end{equation}
and therefore
\begin{equation}
\mathcal A_{kk}
\sim
M_{kk}\ell_k^2
\sim
\ell_k^3.
\label{eq:Akk-scaling}
\end{equation}
A scale-independent characteristic energy per quadratic mode then gives
the dimensional equilibrium prediction
\begin{equation}
\left\langle|\omega_k|^2\right\rangle
\sim
\ell_k^{-3}.
\label{eq:thermal-omega}
\end{equation}

Energy and angular momentum are nonlocal in angular variables, but the argument can be extended.
With $u_k=\ell_k\omega_k$ and
$\mathbf v_k=\dot{\mathbf r}_k$, the exact kinematics gives
\begin{equation}
    \mathbf v_k=\sum_{j=1}^{k}u_j\mathbf t_j,
    \qquad
    \langle u_k^2\rangle
    =\ell_k^2\langle\omega_k^2\rangle
    \propto m_k^{-1},
\end{equation}
where $\mathbf t_j$ is the unit vector perpendicular to link $j$.
Provided equilibrium correlations introduce no additional scale power,
the geometrically growing relative velocities give
\begin{equation}
    \langle|\mathbf v_k|^2\rangle
    \sim\sum_{j=1}^{k}\langle u_j^2\rangle
    \propto\sum_{j=1}^{k}m_j^{-1}
    \propto m_k^{-1}.
\end{equation}
Consequently, the kinetic energy of mass $k$ satisfies
\begin{equation}
    \langle e_k\rangle
    =\frac{m_k}{2}\langle|\mathbf v_k|^2\rangle
    \propto\ell_k^0.
    \label{eq:conservative-equilibrium-energy}
\end{equation}

The angular momentum about the anchor is instead
$j_k=m_k(\mathbf r_k-\mathbf r_0)\wedge\mathbf v_k$.
Since the typical distance from the anchor remains $O(\ell_1)$,
scale-independent geometrical projections and normalized fluctuation
statistics imply
\begin{equation}
    \langle|j_k|\rangle
    \propto m_k\sqrt{\langle|\mathbf v_k|^2\rangle}
    \propto m_k^{1/2}
    \propto\ell_k^{1/2}.
    \label{eq:conservative-equilibrium-angular-momentum}
\end{equation}

These are leading dimensional predictions, rather than an exact equality of the kinetic energies of all masses; the prefactors may vary, for example near the boundaries.
The kinetic energy and angular momentum of the relaxed chain in
Fig.~\ref{fig:noneq_eq}(d) are consistent with these predictions beyond
the first few links.
The thermal tail in Fig.~\ref{fig:crossover} also follows the predicted
angular-velocity scaling, although in this case equipartition is induced
by the Langevin thermostat rather than by conservative relaxation.

\subsection{Equilibrium covariance}

For the stochastic dynamics \eqref{eq:langevin}, the fluctuation--dissipation relation determines the equilibrium statistics
of the thermostat. Since $\mathcal D\mathbf 1=0$, the rigid-rotation mode is
neither dissipated nor stochastically forced. With
$\boldsymbol p=\mathcal A\boldsymbol\omega$, the corresponding conserved
quantity is the total angular momentum
\begin{equation}
J=\mathbf 1^T\boldsymbol p.
\label{eq:J-momentum}
\end{equation}

At fixed configuration, the unconstrained Gibbs covariance is
\begin{equation}
\left\langle
    \boldsymbol{\omega}\boldsymbol{\omega}^T
\right\rangle_{\rm G}
=
\Theta\mathcal A^{-1}.
\end{equation}
For dynamics restricted to a fixed value of $J$, the covariance about the
conditional mean is instead
\begin{equation}
\operatorname{Cov}(\boldsymbol{\omega}\mid J)
=
\Theta
\left[
    \mathcal A^{-1}
    -
    \frac{\mathbf 1\mathbf 1^T}
    {\mathbf 1^T\mathcal A\mathbf 1}
\right].
\label{eq:eq-covariance}
\end{equation}
The second term is a rank-one correction associated with the conserved
rigid-rotation mode and does not modify the ultraviolet scaling.

\section{Ensemble variance scalings}
\label{app:variance-scalings}

Here we summarize the reference scalings shown in
Fig.~\ref{fig:singular}.
We consider both the angular velocities and the canonical momentum
$\boldsymbol p=\mathcal A\boldsymbol\omega$, which is fully nonlocal in the angular variables.
The equilibrium reference is taken from the previous section, condensed into Eq.~\eqref{eq:Akk-scaling}.

\subsection{Short-time stochastic response}

We consider many realizations of the stochastic dynamics \eqref{eq:langevin}, starting from a deterministic initial condition.
All ensemble members start from the same state, so the initial variance
vanishes and its leading growth is determined entirely by the stochastic
increment. In momentum variables,
\begin{equation}
d\boldsymbol p
=
\sqrt{2\eta\Theta}\,
\mathcal G^T d\boldsymbol W
+O(dt),
\end{equation}
with $\mathcal D=\mathcal G^T\mathcal G$. Consequently,
\begin{equation}
\operatorname{Cov}[\boldsymbol p(t)]
=
2\eta\Theta\,t\,\mathcal D
+O(t^2).
\end{equation}
Since $\mathcal D_{kk}\sim\ell_k^{-1}$,
\begin{equation}
\boxed{
\operatorname{Var}[p_k(t)]
\sim
t\,\ell_k^{-1}
}
\qquad (t\rightarrow0).
\label{eq:p-short-scaling}
\end{equation}

For the angular velocities, the same expansion gives
\begin{equation}
\operatorname{Cov}[\boldsymbol\omega(t)]
=
2\eta\Theta\,t\,
\mathcal A^{-1}\mathcal D\mathcal A^{-1}
+O(t^2),
\label{eq:omega-short-cov}
\end{equation}
but because $\mathcal A^{-1}$ is dense and configuration dependent, no simple
power law follows from Eq.~\eqref{eq:omega-short-cov} alone. Numerically, the
early-time data are consistent with an $\ell_k^{-5}$ scaling, which we use as
an empirical reference in Fig.~\ref{fig:singular}.

\subsection{Thermalized scales}

At sufficiently small scales, the ensemble approaches the thermal statistics
set by fluctuation--dissipation balance. The fixed-angular-momentum
covariance in Eq.~\eqref{eq:eq-covariance} introduces a rank-one correction
that does not modify the ultraviolet power-law exponents.
The leading scale dependence may therefore be obtained from
\begin{equation}
\operatorname{Cov}[\boldsymbol p]_{\rm th}
\sim \Theta\mathcal A,
\qquad
\operatorname{Cov}[\boldsymbol\omega]_{\rm th}
\sim \Theta\mathcal A^{-1},
\end{equation}
where the symbol $\sim$ emphasizes that only the scale dependence relevant
here is retained. Using Eq.~\eqref{eq:Akk-scaling} gives
\begin{equation}
\boxed{
\operatorname{Var}[p_k]_{\rm th}
\sim\ell_k^{3},
\qquad
\operatorname{Var}[\omega_k]_{\rm th}
\sim\ell_k^{-3}.
}
\label{eq:thermal-variance-scalings}
\end{equation}
These scalings describe the high-$k$ region once the corresponding degrees of
freedom have relaxed to the fluctuation--dissipation equilibrium.

\subsection{Turbulent saturation}

After the ensemble has lost memory of the common initial condition at scale
$k$, its variance approaches the stationary turbulent variance. From the
stationary cascade scaling measured in Eq.~\eqref{eq:result-omega2},
\begin{equation}
\boxed{
\operatorname{Var}[\omega_k]_{\rm turb}
\sim\ell_k^{-2}.
}
\label{eq:omega-turb-variance}
\end{equation}

For the canonical momentum,
\begin{equation}
\begin{split}
p_k
&=
\sum_{j=1}^N
M_{kj}\ell_k\ell_j
\cos(\theta_j-\theta_k)\omega_j
\\
&=
\sum_{i=k}^N
m_i\ell_k
\sum_{j=1}^i
\cos(\theta_j-\theta_k)u_j .
\end{split}
\label{eq:p-nonlocal-sum}
\end{equation}
Because the outer sum is geometrically dominated by indices $i$ close to
$k$, one has $m_i\ell_k\sim\ell_k^2$ to leading power-law order.
Moreover, the turbulent estimate $u_j=O(1)$ implies that the nonlocal inner
sum introduces no additional algebraic power of $\ell_k$, up to correlations
and finite-range or logarithmic corrections. Hence
$p_k\sim\ell_k^2$ to leading power-law order, giving
\begin{equation}
\boxed{
\operatorname{Var}[p_k]_{\rm turb}
\sim\ell_k^{4}.
}
\label{eq:p-turb-variance}
\end{equation}

The same nonlocal character explains the individual-mass kinetic-energy scaling. From the
definition of the velocity of mass $k$,
\begin{equation}
\dot{\mathbf r}_k
=
\sum_{j=1}^k
u_j\,\mathbf t_j,
\qquad
\mathbf t_j
=
\begin{pmatrix}
    \cos\theta_j\\
    \sin\theta_j
\end{pmatrix},
\end{equation}
the kinetic energy carried by that mass is
\begin{equation}
e_k
=
\frac{m_k}{2}
\left|
    \sum_{j=1}^k u_j\mathbf t_j
\right|^2
=
\frac{m_k}{2}
\sum_{i,j=1}^k
u_i u_j
\cos(\theta_i-\theta_j).
\label{eq:energy-nonlocal-sum}
\end{equation}
Again using the turbulent dimensional scaling
$u_j=O(1)$, the sum introduces no
additional algebraic power of $\ell_k$ to leading order. Since
$m_k\sim\ell_k$, one therefore obtains
\begin{equation}
\boxed{
\langle e_k\rangle_{\rm turb}
\sim\ell_k.
}
\label{eq:energy-turb-scaling}
\end{equation}
The same argument applies to the angular momentum $j_k=m_k(\mathbf r_k-\mathbf r_0)\wedge\mathbf v_k$. Since the distance from the anchor remains $O(\ell_1)$ and the velocity introduces no additional power of $\ell_k$, we obtain $\langle|j_k|\rangle_{\rm turb}\sim m_k\sim\ell_k$, assuming no additional scale dependence from geometrical correlations.

Thus the canonical momentum and kinetic energy can retain their
nonequilibrium turbulent scalings even when the ultraviolet angular-velocity
modes have crossed over to the thermal
$\langle|\omega_k|^2\rangle\sim\ell_k^{-3}$ spectrum. This reflects their
intrinsically nonlocal dependence on larger-scale degrees of freedom.
The conservative predictions in Appendix~\ref{app:equipartition} concern
equilibrium across the hierarchy; in the driven state, thermalization
of small-scale angular velocities alone need not produce the same
scalings for these cumulative observables over the scales studied.

The resulting reference regimes for the ensemble variance are
\begin{equation}
\begin{array}{c|ccc}
 & \text{short time}
 & \text{thermal}
 & \text{turbulent}
\\[2pt] \hline\\[-8pt]
\operatorname{Var}[\omega_k]
 & \ell_k^{-5}\;(\text{empirical})
 & \ell_k^{-3}
 & \ell_k^{-2}
\\[4pt]
\operatorname{Var}[p_k]
 & \ell_k^{-1}
 & \ell_k^{3}
 & \ell_k^{4}
\end{array}
\end{equation}
The moving crossovers between these regimes constitute the variance front
observed in Fig.~\ref{fig:singular}.

\section{Velocity-increment regularity and a tunable dimensional critical exponent}
\label{sec:onsager_holder}

The anomalous dissipation observed in the main text motivates the question
of which regularity conditions ensure energy conservation in the infinite
inviscid chain. We relate velocity increments to the mechanical kinematics
and show how independent mass and length ratios tune the dimensional
constant-flux exponent. This provides a regularity interpretation of the
cascade, analogous in spirit to Onsager's argument \cite{onsager1949},
under the comparable-scale transfer assumption used below. The
infinite-chain discussion concerns an unforced, inviscid hierarchy with a
fixed anchor.

\subsection{Geometric hierarchy and velocity increments}

Consider a hierarchy in which lengths and masses decrease at independent
geometric rates,
\begin{equation}
    \ell_k=\ell_1r_\ell^{k-1},\quad
    m_k=m_1r_m^{k-1},\qquad
    0<r_\ell,r_m<1.
    \label{eq:generalized-hierarchy}
\end{equation}
Introducing the dimensionless scale
$q_k=\ell_k/\ell_1=r_\ell^{k-1}$, we may write
\begin{equation}
    m_k=m_1 q_k^\alpha,\qquad
    \alpha=\frac{\log r_m}{\log r_\ell}>0.
    \label{eq:mass-length-exponent}
\end{equation}
Thus $\alpha$ characterizes the relative decrease of mass and length.
The hierarchy used in the main text corresponds to
$r_\ell=r_m=1/2$ and $\alpha=1$.

Both total length and total mass are finite. Since the sum of all link
lengths is finite, the positions $\mathbf r_k$ approach a limiting position
$\mathbf r_*$ as $k\to\infty$. This limiting position is the accumulation
point of the chain: it is approached by infinitely many masses connected by
progressively shorter links. The remaining arclength $s_k$ bounds the
distance to this point:
\begin{equation}
    |\mathbf r_*-\mathbf r_k|
    \leq s_k,\qquad
    s_k=\sum_{j=k+1}^{\infty}\ell_j
    =\frac{\ell_{k+1}}{1-r_\ell}.
    \label{eq:tail-arclength}
\end{equation}

The mechanical kinematics provides a natural measure of regularity.
Writing $\mathbf v_k=\dot{\mathbf r}_k$ and
$\mathbf t_k=(\cos\theta_k,\sin\theta_k)^T$, the unit vector perpendicular
to link $k$, we have
\begin{equation}
    \mathbf v_k-\mathbf v_{k-1}=u_k\mathbf t_k,\quad
    u_k=\ell_k\omega_k,\quad
    |\mathbf v_k-\mathbf v_{k-1}|=|u_k|.
    \label{eq:velocity_increment}
\end{equation}
The magnitude of the velocity increment across a link is therefore $|u_k|$.
We consider the discrete regularity bound
\begin{equation}
    |u_k(t)|\leq U(t)q_k^h\quad\hbox{for all }k,
    \label{eq:holder_scaling}
\end{equation}
where $U(t)$ is independent of $k$. For $h>0$, summing the increments
between links $k$ and $K>k$ gives
\begin{equation*}
    |\mathbf v_K-\mathbf v_k|
    \leq\sum_{j=k+1}^{K}|u_j|
    \leq U(t)\sum_{j=k+1}^{K}q_j^h
    \leq\frac{U(t)q_{k+1}^h}{1-r_\ell^h}.
\end{equation*}
The geometric-series bound tends to zero as $k\to\infty$, independently of
$K$. Hence the velocities approach a limit $\mathbf v_*(t)$, with
\begin{equation}
    |\mathbf v_*(t)-\mathbf v_k(t)|
    \leq \frac{U(t)q_{k+1}^{h}}{1-r_\ell^h}
    =\frac{U(t)}{1-r_\ell^h}
      \left(\frac{(1-r_\ell)s_k}{\ell_1}\right)^h.
    \label{eq:holder-tail-bound}
\end{equation}
For $0<h\leq1$, this is a H\"older-type bound in the distance $s_k$ measured
along the chain to its limiting endpoint. Two masses on a folded chain may
lie close together in the plane while being far apart along the rods. The
bound therefore describes regularity along the chain, without assuming a
corresponding velocity field in the surrounding plane.

At $h=0$, the bound allows increments that remain finite as the links become
shorter, but also increments that decrease toward zero. It therefore does
not determine whether the velocities converge. If $|u_k|$ stays above a
fixed positive value for arbitrarily large $k$, convergence is impossible.
For $h<0$, the upper bound grows toward small scales, allowing singular
increments. Actual scaling $|u_k|\sim q_k^h$ would imply divergent
increments; the upper bound alone does not. Negative $h$ is used here to
describe singular scaling rather than H\"older continuity.

\subsection{Dimensional transfer estimate and the candidate exponent}

As in Sec.~\ref{sec:scaling-prediction}, we estimate the transfer by assuming
that it is controlled predominantly by motions at comparable scales. The
characteristic energy $m_k|u_k|^2$ and transfer time
$t_k^{\rm tr}\sim\ell_k/|u_k|$ then give
\begin{equation}
    \Pi_k^{\rm loc}
    \sim\frac{m_k|u_k|^2}{t_k^{\rm tr}}
    \sim m_k\frac{|u_k|^3}{\ell_k}.
    \label{eq:holder_flux_estimate}
\end{equation}
Here $\Pi_k^{\rm loc}$ denotes a local estimate of the transfer magnitude,
distinguished from the signed exact flux $\Pi_k$ in
Eq.~\eqref{eq:local-budget}.

For $|u_k|\sim U_0 q_k^h$, with $U_0$ independent of scale, the mass,
velocity, and inverse-length factors yield
\begin{equation}
    \Pi_k^{\rm loc}\sim
    \frac{m_1U_0^3}{\ell_1}
    q_k^\alpha q_k^{3h}q_k^{-1}
    =\frac{m_1U_0^3}{\ell_1}q_k^{\alpha+3h-1}.
    \label{eq:holder_flux_scaling}
\end{equation}
A finite nonzero constant flux in this estimate requires
$\alpha+3h-1=0$. The resulting candidate exponent $h_c$ and the associated
velocity scalings are
\begin{equation}
    \boxed{h_c=\frac{1-\alpha}{3}},\qquad
    |u_k|\sim q_k^{h_c},\qquad
    |\omega_k|\sim q_k^{-(\alpha+2)/3}.
    \label{eq:critical_holder_general}
\end{equation}
For the hierarchy studied in the main text,
\begin{equation}
    \boxed{h_c=0\quad(\alpha=1)},
    \label{eq:critical_holder}
\end{equation}
which gives $|\omega_k|\sim q_k^{-1}$ and recovers
Eq.~\eqref{eq:kolmogorov-scaling}. For $h>h_c$, the positive exponent in
Eq.~\eqref{eq:holder_flux_scaling} instead gives
\begin{equation}
    h>h_c\quad\Longrightarrow\quad\Pi_k^{\rm loc}\to0.
    \label{eq:holder_energy_conservation}
\end{equation}
This dimensional conclusion becomes an energy-conservation criterion only
if an appropriate bound on the exact flux can be established. At $h=h_c$,
the estimate permits a nonzero flux, while signs, correlations, and
cancellations determine the actual mean transfer. Likewise, $h<h_c$ alone
does not imply anomalous dissipation.

Independent mass and length ratios thus allow the dimensional exponent to
be varied: $0<\alpha<1$ gives $0<h_c<1/3$, $\alpha=1$ gives $h_c=0$, and
$\alpha>1$ gives $h_c<0$. The last case corresponds to a candidate singular
increment scaling rather than classical H\"older regularity.

The decreasing masses can make singular increments compatible with finite
total kinetic energy. For $h<0$, summing the increments from the fixed
anchor gives $|\mathbf v_k|\leq C_h U q_k^h$, with $C_h$ independent of $k$.
Consequently,
\begin{equation*}
    \sum_k m_k|\mathbf v_k|^2
    \leq m_1 C_h^2 U^2\sum_k q_k^{\alpha+2h}
    <\infty\qquad\hbox{if }\alpha+2h>0.
\end{equation*}
At the dimensional value $h_c$, the exponent in this sum is
$\alpha+2h_c=(\alpha+2)/3>0$. A negative $h_c$ is therefore compatible with
finite energy.

For comparison, the dimensional energy flux per unit mass in Euler
turbulence is $\delta v(r)^3/r\sim r^{3h-1}$, giving the familiar exponent
$1/3$ \cite{onsager1949}. In the chain, the observable is a total mechanical
power, and the scale-dependent mass adds $\alpha$ to the exponent. For
$\alpha=1$ it compensates the inverse length, giving $h_c=0$. The analogy
concerns the dimensional transfer estimate. Establishing an Onsager-type
energy-conservation criterion would require control of the exact nonlocal
mechanical flux in terms of velocity increments, with suitable time
integrability. Such a result is not established here, and $h_c$ remains a
candidate threshold from constant-flux phenomenology.

\section{Visualizations: supplemental movies and interactive simulation}
\label{app:movies}

Three movies, available as ancillary files of this submission \cite{supplemental}, are intended for visualization purposes. In addition, an interactive simulation of the chain runs online \cite{website}. The quantitative results are presented in the figures of the main text.

\emph{Movie S1: the turbulent chain} (\texttt{Movie\_S1.mp4}, 30\,s). A
single trajectory of the deterministic, driven-dissipative chain in its
statistically stationary state. The pivot and the first link are drawn in
black and the remaining links colored by their index $k$. The thin line
traces the recent path of the free end of the chain and fades with time,
showing the erratic small-scale motion that coexists with the slower
rotation of the large links.
The chain, with $20$ links, has a scale ratio $\gamma=2$ and viscosity is $\eta=10^{-13}$.

\emph{Movie S2: upscale propagation of microscopic uncertainty}
(\texttt{Movie\_S2.mp4}, 75\,s). The $128$ realizations of
Fig.~\ref{fig:sptsto_viz}, evolving from an identical initial condition
under independent thermal noise, for $\eta=10^{-13}$
(Sec.~\ref{sec:results-spst}). The time, shown above the chain, runs from
$t/\tau_0=1.4\times10^{-6}$ to $t/\tau_0=1.7\times10^{-2}$. Realizations are
drawn semi-transparently, so that regions where they coincide appear dark
and the spread of the ensemble appears as a halo. The halo forms at the
free end and climbs toward the pivot while the large links remain
indistinguishable across realizations. In each frame the configurations
are rotated so that the mean position of the free end lies vertically
below the pivot. For visual clarity, the lengths and masses decrease by a
factor $1.3$ per link rather than $2$.

\emph{Movie S3: a branching hierarchy} (\texttt{Movie\_S3.mp4}, 30\,s). Simulation of the branching extension discussed in
Sec.~\ref{sec:discussion}: a tree of V-shaped rigid rotors in which each
rotor carries, at the ends of its arms, two copies of itself scaled down by
a constant factor. The root rotor is driven by shaking its pivot along the
horizontal line, which exerts a torque on every body, and the thin line
traces the recent path of one of the outer rotors. This system is not
analysed in the present paper.
The hierarchy has $10$ levels with a scale ratio $\gamma=2.5$, and the angle of the V-shaped rotors is $115^\circ$. As in the chain, dissipation is more effective at smaller scales and acts on relative motion, thus modeling joint friction. The forcing is also different: energy is injected through the boundary conditions by oscillating the pivot of the root rotor. This is equivalent to an oscillating gravitational field, resulting in a power-law forcing across the hierarchy.

\emph{Interactive simulation} \cite{website}. A web page runs the
chain live in the browser, with no installation. Its integrator is a port of
the code used for the simulations of this paper, with the same implicit--explicit
scheme and $\gamma=2$. The number of links, the
viscosity, the forcing (periodic, constant-power or none), its amplitude and
frequency, and an optional gravitational field can be changed during the
run, individual links can be flicked with the pointer, and the
time-averaged kinetic energy of each link is displayed as the chain evolves.
The page is meant for exploration; all quantitative results of this paper
come from the production code.

\bibliography{apssamp}

\end{document}